\documentclass[trackchanges,twocolumn]{aastex701}
\usepackage{natbib}
\usepackage{graphicx}
\usepackage{gensymb}
\usepackage[english]{babel}
\usepackage{wrapfig}
\usepackage{appendix}
\usepackage{float}
\usepackage{booktabs}
\usepackage{tabularx}
\usepackage{multirow}
\usepackage{amssymb}
\usepackage{amsmath}
\usepackage{verbatim}
\usepackage{xspace}
\usepackage{listings}
\usepackage{hyperref}

\graphicspath{{./}}

\begin{document}

\newcommand{\jten}{J1012$+$5307}
\newcommand{\jzerosix}{J0645$+$5158}
\newcommand{\jtwentyone}{J2145$-$0750}
\newcommand{\jtwentythree}{J2302$+$4442}

\title{CHIME-o-Grav: Wideband Timing of Four Millisecond Pulsars from the NANOGrav 15-yr dataset}
%% Current working title until official one is determined

\author[0000-0001-5134-3925]{Gabriella Y. Agazie}
\affiliation{Center for Gravitation, Cosmology and Astrophysics, Department of Physics and Astronomy, University of Wisconsin-Milwaukee,\\ P.O. Box 413, Milwaukee, WI 53201, USA}
\email{}
\author[0000-0001-6295-2881]{David L. Kaplan}
\affiliation{Center for Gravitation, Cosmology and Astrophysics, Department of Physics and Astronomy, University of Wisconsin-Milwaukee,\\ P.O. Box 413, Milwaukee, WI 53201, USA}
\email{}
\author[0000-0002-2820-0931]{Abhimanyu Susobhanan}
\affiliation{Max-Planck-Institut f{\"u}r Gravitationsphysik (Albert-Einstein-Institut), Callinstra{\ss}e 38, D-30167 Hannover, Germany\\Leibniz Universit{\"a}t Hannover, D-30167 Hannover, Germany}
\email{}
\author[0000-0001-9784-8670]{Ingrid H. Stairs}
\affiliation{Department of Physics and Astronomy, University of British Columbia, 6224 Agricultural Road, Vancouver, BC V6T 1Z1, Canada}
\email{}
\author[0000-0003-1884-348X]{Deborah C. Good}
\affiliation{Department of Physics and Astronomy, University of Montana, 32 Campus Drive, Missoula, MT 59812}
\email{}
\author[0000-0001-8845-1225]{Bradley W. Meyers}
\affiliation{Australian SKA Regional Centre (AusSRC), Curtin University, Bentley, WA 6102, Australia}
\affiliation{International Centre for Radio Astronomy Research (ICRAR), Curtin University, Bentley, WA 6102, Australia}
\email{}
\author[0000-0001-8384-5049]{Emmanuel Fonseca}
\affiliation{Department of Physics and Astronomy, West Virginia University, P.O. Box 6315, Morgantown, WV 26506, USA}
\affiliation{Center for Gravitational Waves and Cosmology, West Virginia University, Chestnut Ridge Research Building, Morgantown, WV 26505, USA}
\email{}
\author[0000-0001-5465-2889]{Timothy T. Pennucci}
\affiliation{Institute of Physics and Astronomy, E\"{o}tv\"{o}s Lor\'{a}nd University, P\'{a}zm\'{a}ny P. s. 1/A, 1117 Budapest, Hungary}
\email{}
\author[0000-0002-8935-9882]{Akash Anumarlapudi}
\affiliation{Department of Physics and Astronomy, University of North Carolina, Chapel Hill, NC 27599, USA}
\email{}
\author[0000-0003-0638-3340]{Anne M. Archibald}
\affiliation{Newcastle University, NE1 7RU, UK}
\email{}
\author[0009-0008-6187-8753]{Zaven Arzoumanian}
\affiliation{X-Ray Astrophysics Laboratory, NASA Goddard Space Flight Center, Code 662, Greenbelt, MD 20771, USA}
\email{}
\author[0000-0003-2745-753X]{Paul T. Baker}
\affiliation{Department of Physics and Astronomy, Widener University, One University Place, Chester, PA 19013, USA}
\email{}
\author[0000-0003-3053-6538]{Paul R. Brook}
\affiliation{Institute for Gravitational Wave Astronomy and School of Physics and Astronomy, University of Birmingham, Edgbaston, Birmingham B15 2TT, UK}
\email{}
\author[0009-0007-0757-9800]{Alyssa Cassity}
\affiliation{Department of Physics and Astronomy, University of British Columbia, 6224 Agricultural Road, Vancouver, BC V6T 1Z1, Canada}
\email{}
\author[0000-0002-6039-692X]{H. Thankful Cromartie}
\affiliation{National Research Council Research Associate, National Academy of Sciences, Washington, DC 20001, USA resident at Naval Research Laboratory, Washington, DC 20375, USA}
\email{}
\author[0000-0002-1529-5169]{Kathryn Crowter}
\affiliation{Department of Physics and Astronomy, University of British Columbia, 6224 Agricultural Road, Vancouver, BC V6T 1Z1, Canada}
\email{}
\author[0000-0002-2185-1790]{Megan E. DeCesar}
\altaffiliation{Resident at the Naval Research Laboratory}
\affiliation{Department of Physics and Astronomy, George Mason University, Fairfax, VA 22030, resident at the U.S. Naval Research Laboratory, Washington, DC 20375, USA}
\email{}
\author[0000-0002-6664-965X]{Paul B. Demorest}
\affiliation{National Radio Astronomy Observatory, 1003 Lopezville Rd., Socorro, NM 87801, USA}
\email{}
\author[0000-0001-8885-6388]{Timothy Dolch}
\affiliation{Department of Physics, Hillsdale College, 33 E. College Street, Hillsdale, MI 49242, USA}
\affiliation{Eureka Scientific, 2452 Delmer Street, Suite 100, Oakland, CA 94602-3017, USA}
\email{}
\author[0000-0003-4098-5222]{Fengqiu Adam Dong}
\affiliation{National Radio Astronomy Observatory, 520 Edgemont Road, Charlottesville, VA 22903, USA}
\email{}
\author[0000-0001-7828-7708]{Elizabeth C. Ferrara}
\affiliation{Department of Astronomy, University of Maryland, College Park, MD 20742, USA}
\affiliation{Center for Research and Exploration in Space Science and Technology, NASA/GSFC, Greenbelt, MD 20771}
\affiliation{NASA Goddard Space Flight Center, Greenbelt, MD 20771, USA}
\email{}
\author[0000-0001-5645-5336]{William Fiore}
\affiliation{Department of Physics and Astronomy, University of British Columbia, 6224 Agricultural Road, Vancouver, BC V6T 1Z1, Canada}
\email{}
\author[0000-0001-7624-4616]{Gabriel E. Freedman}
\affiliation{NASA Goddard Space Flight Center, Greenbelt, MD 20771, USA}
\email{}
\author[0000-0001-6166-9646]{Nate Garver-Daniels}
\affiliation{Department of Physics and Astronomy, West Virginia University, P.O. Box 6315, Morgantown, WV 26506, USA}
\affiliation{Center for Gravitational Waves and Cosmology, West Virginia University, Chestnut Ridge Research Building, Morgantown, WV 26505, USA}
\email{}
\author[0000-0001-8158-683X]{Peter A. Gentile}
\affiliation{Department of Physics and Astronomy, West Virginia University, P.O. Box 6315, Morgantown, WV 26506, USA}
\affiliation{Center for Gravitational Waves and Cosmology, West Virginia University, Chestnut Ridge Research Building, Morgantown, WV 26505, USA}
\email{}
\author[0000-0003-4090-9780]{Joseph Glaser}
\affiliation{Department of Physics and Astronomy, West Virginia University, P.O. Box 6315, Morgantown, WV 26506, USA}
\affiliation{Center for Gravitational Waves and Cosmology, West Virginia University, Chestnut Ridge Research Building, Morgantown, WV 26505, USA}
\email{}
\author[0000-0003-2742-3321]{Jeffrey S. Hazboun}
\affiliation{Department of Physics, Oregon State University, Corvallis, OR 97331, USA}
\email{}
\author[0000-0003-1082-2342]{Ross J. Jennings}
\altaffiliation{NANOGrav Physics Frontiers Center Postdoctoral Fellow}
\affiliation{Department of Physics and Astronomy, West Virginia University, P.O. Box 6315, Morgantown, WV 26506, USA}
\affiliation{Center for Gravitational Waves and Cosmology, West Virginia University, Chestnut Ridge Research Building, Morgantown, WV 26505, USA}
\email{}
\author[0000-0001-6607-3710]{Megan L. Jones}
\affiliation{Center for Gravitation, Cosmology and Astrophysics, Department of Physics and Astronomy, University of Wisconsin-Milwaukee,\\ P.O. Box 413, Milwaukee, WI 53201, USA}
\email{}
\author[0000-0002-0893-4073]{Matthew Kerr}
\affiliation{Space Science Division, Naval Research Laboratory, Washington, DC 20375-5352, USA}
\email{}
\author[0000-0003-0721-651X]{Michael T. Lam}
\affiliation{SETI Institute, 339 N Bernardo Ave Suite 200, Mountain View, CA 94043, USA}
\affiliation{School of Physics and Astronomy, Rochester Institute of Technology, Rochester, NY 14623, USA}
\affiliation{Laboratory for Multiwavelength Astrophysics, Rochester Institute of Technology, Rochester, NY 14623, USA}
\email{}
\author[0000-0003-1301-966X]{Duncan R. Lorimer}
\affiliation{Department of Physics and Astronomy, West Virginia University, P.O. Box 6315, Morgantown, WV 26506, USA}
\affiliation{Center for Gravitational Waves and Cosmology, West Virginia University, Chestnut Ridge Research Building, Morgantown, WV 26505, USA}
\email{}
\author[0000-0001-5373-5914]{Jing Luo}
\altaffiliation{Deceased}
\affiliation{Department of Astronomy \& Astrophysics, University of Toronto, 50 Saint George Street, Toronto, ON M5S 3H4, Canada}
\email{}
\author[0000-0001-5229-7430]{Ryan S. Lynch}
\affiliation{Green Bank Observatory, P.O. Box 2, Green Bank, WV 24944, USA}
\email{}
\author[0000-0001-5481-7559]{Alexander McEwen}
\affiliation{Center for Gravitation, Cosmology and Astrophysics, Department of Physics and Astronomy, University of Wisconsin-Milwaukee,\\ P.O. Box 413, Milwaukee, WI 53201, USA}
\email{}
\author[0000-0002-2885-8485]{James W. McKee}
\affiliation{Department of Physics and Astronomy, Union College, Schenectady, NY 12308, USA}
\email{}
\author[0000-0001-7697-7422]{Maura A. McLaughlin}
\affiliation{Department of Physics and Astronomy, West Virginia University, P.O. Box 6315, Morgantown, WV 26506, USA}
\affiliation{Center for Gravitational Waves and Cosmology, West Virginia University, Chestnut Ridge Research Building, Morgantown, WV 26505, USA}
\email{}
\author[0000-0002-4642-1260]{Natasha McMann}
\affiliation{Department of Physics and Astronomy, Vanderbilt University, 2301 Vanderbilt Place, Nashville, TN 37235, USA}
\email{}
\author[0000-0002-3616-5160]{Cherry Ng}
\affiliation{Dunlap Institute for Astronomy and Astrophysics, University of Toronto, 50 St. George St., Toronto, ON M5S 3H4, Canada}
\email{}
\author[0000-0002-6709-2566]{David J. Nice}
\affiliation{Department of Physics, Lafayette College, Easton, PA 18042, USA}
\email{}
\author[0000-0002-8509-5947]{Benetge B. P. Perera}
\affiliation{Arecibo Observatory, HC3 Box 53995, Arecibo, PR 00612, USA}
\email{}
\author[0000-0002-8826-1285]{Nihan S. Pol}
\affiliation{Department of Physics, Texas Tech University, Box 41051, Lubbock, TX 79409, USA}
\email{}
\author[0000-0002-2074-4360]{Henri A. Radovan}
\affiliation{Department of Physics, University of Puerto Rico, Mayag\"{u}ez, PR 00681, USA}
\email{}
\author[0000-0001-5799-9714]{Scott M. Ransom}
\affiliation{National Radio Astronomy Observatory, 520 Edgemont Road, Charlottesville, VA 22903, USA}
\email{}
\author[0000-0002-5297-5278]{Paul S. Ray}
\affiliation{Space Science Division, Naval Research Laboratory, Washington, DC 20375-5352, USA}
\email{}
\author[0000-0001-7832-9066]{Alexander Saffer}
\altaffiliation{NANOGrav Physics Frontiers Center Postdoctoral Fellow}
\affiliation{National Radio Astronomy Observatory, 520 Edgemont Road, Charlottesville, VA 22903, USA}
\email{}
\author[0000-0003-4391-936X]{Ann Schmiedekamp}
\affiliation{Department of Physics, Penn State Abington, Abington, PA 19001, USA}
\email{}
\author[0000-0002-1283-2184]{Carl Schmiedekamp}
\affiliation{Department of Physics, Penn State Abington, Abington, PA 19001, USA}
\email{}
\author[0000-0002-7283-1124]{Brent J. Shapiro-Albert}
\affiliation{Department of Physics and Astronomy, West Virginia University, P.O. Box 6315, Morgantown, WV 26506, USA}
\affiliation{Center for Gravitational Waves and Cosmology, West Virginia University, Chestnut Ridge Research Building, Morgantown, WV 26505, USA}
\affiliation{Giant Army, 915A 17th Ave, Seattle WA 98122}
\email{}
\author[0000-0002-7261-594X]{Kevin Stovall}
\affiliation{National Radio Astronomy Observatory, 1003 Lopezville Rd., Socorro, NM 87801, USA}
\email{}
\author[0000-0002-1075-3837]{Joseph K. Swiggum}
\altaffiliation{NANOGrav Physics Frontiers Center Postdoctoral Fellow}
\affiliation{Department of Physics, Lafayette College, Easton, PA 18042, USA}
\email{}
\author[0009-0001-5938-5000]{Mercedes S. Thompson}
\affiliation{Department of Physics and Astronomy, University of British Columbia, 6224 Agricultural Road, Vancouver, BC V6T 1Z1, Canada}
\email{}
\author[0000-0001-9678-0299]{Haley M. Wahl}
\affiliation{Department of Physics and Astronomy, West Virginia University, P.O. Box 6315, Morgantown, WV 26506, USA}
\affiliation{Center for Gravitational Waves and Cosmology, West Virginia University, Chestnut Ridge Research Building, Morgantown, WV 26505, USA}
\email{}

\noaffiliation

\correspondingauthor{Gabriella Y. Agazie}
\email{gabriella.agazie@nanograv.org}

%Tiered author list with major contributors first then everyone else alphabetical
%NANOGrav authors: According to NANOGrav authorship policy this paper should include major contributors + 15yr timers but not all full members as this does not include any GW results or limits. 

%CHIME authors: According to MOU and CHIME/Pulsar authorship policy, people can opt-into authorship when the paper is first circulated. 

\begin{abstract}

Wideband timing of the the North American Nanohertz Observatory for Gravitational Waves (NANOGrav) datasets was first done for the 12.5 yr dataset. This method, where a single time-of-arrival (TOA) and a single dispersion measure (DM) are measured using the entire bandwidth of each observation, proved to be invaluable for characterizing the time-varying dispersion measure, improving handling of frequency dependent profile variability, and data volume reduction.
The Canadian Hydrogen Intensity Mapping Experiment (CHIME) Telescope has been observing most NANOGrav millisecond pulsars (MSPs) at nearly daily cadence (compared to roughly monthly cadence for other NANOGrav observations) since 2019 with the objective of integration into future pulsar timing array (PTA) datasets. In this paper, we show the results of integration of high-cadence, low-observing-frequency CHIME data with data from the NANOGrav experiment for an isolated MSP PSR~\jzerosix\ and three binary MSPs PSR~\jten, PSR~\jtwentyone, and PSR~\jtwentythree. Using a wideband timing pipeline which we also describe, we present updated timing results for all four sources, including improved relativistic post-Keplerian measurements for the three binary pulsars in this analysis. For PSR~\jtwentythree, we report an updated strong detection of Shapiro delay from which we measured a companion mass of $0.35^{+0.05}_{-0.04}\ M_{\odot}$, a pulsar mass of $1.8^{+0.3}_{-0.3}\ M_{\odot}$, and an orbital inclination of ${80\degree}^{+1}_{-2}$. We also report updated constraints on the reflex motion for PSR~\jtwentyone\ using a combination of Very Long Baseline Array astrometry and our updated measurement of the time derivative of the projected semi-major axis of the pulsar orbit as a prior. 

%Ending sentence 

\end{abstract}

% \keywords{}

\section{Introduction \label{sec:intro}}

Pulsars are a class of highly magnetized neutron stars characterized by beams of electromagnetic radiation emanating from each magnetic pole which terrestrial observers detect as a pulsed emission \citep{msp_living_review}. They have densities comparable to atomic nuclei and are useful laboratories for studying extreme states of matter, placing constraints on the nuclear equation of state, and testing theories of gravity \citep{nuclear_eos,pulsar_handbook}. 

Millisecond pulsars (MSPs) are a class of pulsars that have very low spin periods ($P \lesssim 30$ ms) and low period derivatives ($\dot{P} \lesssim 10^{-19} \text{ s s}^{-1}$). Many MSPs form in binaries where the kick from the explosion of the supernovae of the pulsar progenitor does not disrupt the binary \citep{1991_Bhattacharya}. Over time matter from the companion is accreted by the pulsar by Roche Lobe overflow and the corresponding transfer of angular momentum speeds up the pulsar's rotation to millisecond spin periods \citep{1982_Alpig}. These systems typically become white dwarf-MSP binaries with nearly circular orbits\citep{1992_Phinney}. Pulsar timing arrays (PTAs) are constructed from long term timing campaigns of MSPs and are used to search for nanohertz frequency gravitational waves which manifest as correlated perturbations in pulse times-of-arrival (TOA) residuals from different pulsars \citep{1983_Hellings_Downs,NG15_GWB,2023_epta_gwb,2025_mpta_gwb,2023_ppta_gwb,2023_cpta_gwb}.

%They are a class of pulsars with very low spin periods (<30ms) and low period derivatives ($<10^{-20} \text{ s s}^{-1}$)

%How do they form : When one of the stars in a binary system that is massive enough to become a neutron star undergoes supernovae. If the kick from the explosion does not disrupt the binary, the new pulsar might  start accreting matter from the comopanion star through Roche Lobe overflow and the transfer of angular momentum eventually causes the pulsar to be sped up to MSP rotational periods (recycling )

High-precision timing of binary MSPs can also be used to probe parameters of the nuclear equation of state such as neutron star masses, radii, and system inclinations \citep{2016_Ozel}. In pulsar binaries, we can measure relativistic corrections to the Keplerian orbital model which are represented in pulsar ephemerides as post-Keplerian (PK) parameters. Measurement of a single PK parameter, such as the relativistic orbital decay ($\dot{P}_{B}$), can place constraints on possible combinations of pulsar and companion masses by defining a curve in companion-pulsar mass space \citep{1982_taylor_weisberg}. Shapiro delay, which is caused by pulsar signal passing through the curved spacetime around companion star, is typically measured in nearly edge-on orbits and is parameterized by two PK parameters, range ($r$) and shape ($s$) \citep{1964_Shapiro, dd_model_2}. These parameters can be used to get measurements of the pulsar and companion masses as well as the system inclination \citep{ELL1H_model}.  Having three or more measured PK parameters makes the system over-determined, allowing it to be used to test theories of gravity \citep{nuclear_eos}. 

%Specifically mention how Pbdot and xdot put constraints on the pulsar and companion masses. xdot specifically puts constraints on reflex motion

%Pulsar astrometry is typically difficult to accurately measure via pulsar timing methods alone 

\subsection{Wideband Timing} \label{subseb:wb_timing}

TOAs are usually calculated by cross-correlating observed pulse profiles with a noise-free template of the integrated pulse profile \citep{pulsar_handbook}. In traditional narrowband timing this template is frequency-averaged and does not account for evolution of the pulse profile across the observing bandwidth. When timing pulsars, we can expect some evolution of the pulse profile shape as a function of frequency caused by intrinsic changes in the pulsar magnetosphere, and extrinsic effects such as interstellar scattering and frequency-dependent dispersion measure (DM) \citep{1986_hankins,12.5yr_wideband}. Narrowband timing methods typically deal with profile evolution by calculating TOAs at several different frequencies in a single observation and the inclusion of arbitrary frequency dependent parameters in the timing model \citep{12.5_year_narrowband}.

Wideband (WB) timing is a method of using a smooth 2-D frequency-dependent model of the pulse profile, often called a template portrait, to simultaneously measure a TOA and a dispersion measure (DM) from each observation, without splitting the observations into multiple frequency sub-bands \citep{2014_pennucci_demorest}.
The reference frequency for the TOA is selected such that there is zero covariance between the TOA phase offset and DM measurement. WB template portraits are calculated from high signal to noise (S/N) data portraits, which are constructed by aligning and averaging many observations with strong detections of the pulsar together \citep{pulse_portraiture}.

To construct the template portrait, the mean (frequency-averaged) profile-subtracted data portrait is decomposed into an orthonormal set of eigenprofiles using principal component analysis (PCA). A small number of the eigenprofiles are kept as the basis for constructing the profile evolution model. The eigenprofiles and the mean profile are then de-noised using a Stationary Wavelet Transform \citep{2019_lee}. Coordinate functions of the profile evolution are constructed by projecting the de-noised eigenprofiles onto the de-noised mean profile. These are then approximated by fitting a spline to each coordinate function. The coordinate function splines and de-noised mean profile are used to construct the 2-D template portrait which is then used for calculation WB TOAs.

Calculating TOAs with this method means that individual wideband TOAs often have much greater precision than individual narrowband TOAs derived from the same data. Only one wideband TOA measurement need be made from a single observation to cover the total observing bandwidth, whereas multiple narrowband TOAs are needed to cover the total bandwidth and then are fit simultaneously, with a final precision comparable to that of the wideband data. However, using multiple narrowband TOAs may require arbitrary frequency dependent parameters in the timing model because profile evolution is handled at the timing stage rather than the profile template creation stage. 
The use of wideband timing significantly reduces the TOA volume in large PTA datasets, lowers the number of free parameters in the timing model, and more smoothly handles frequency dependent profile variability when compared to narrowband timing \citep{NG15_gwb_methods,NG15_timing,12.5yr_wideband}.

\subsection{The Canadian Hydrogen Intensity Mapping Experiment}

The Canadian Hydrogen Intensity Mapping Experiment (CHIME) Telescope is a drift scan radio telescope that has been observing sources in the NANOGrav pulsar timing array  with nearly daily cadence since 2019 \citep{chime_overview,Good_2021}. In this paper, we present results of wideband data combination of CHIME data with that from the NANOGrav 15 year dataset (NG15) for four pulsars: PSR~\jzerosix, PSR~\jten, PSR~\jtwentyone, and PSR~\jtwentythree. In Section~\ref{sec:observations} we describe the instruments used for observing and the data reduction techniques used to calculate the CHIME WB TOAs. Section~\ref{sec:timing_analysis} describes the binary models, DM modeling, and noise modeling used for the timing analysis of the combined CHIME/NG15 dataset. Section~\ref{sec:results} reports timing analysis results for each source, and Section~\ref{sec:discussion} discusses the impact of the inclusion of CHIME data of our timing results. Finally, Section~\ref{sec:conclusions} summarizes our major conclusions and areas for future work.

\section{Observations and Data Reduction} \label{sec:observations}

The observations for all four sources in this analysis were conducted on both the Robert C. Byrd Green Bank Telescope (GBT) and CHIME telescopes. All GBT data used in this analysis came from the NANOGrav observing program as described in \citet{NG15_timing}. The total observing time span for GBT data varies for each source based on when each was included in the observing program and is listed in Table~\ref{tab:timing_stats}.

\subsection{GBT}\label{subsec:nano_obs}

For each observing epoch, the GBT data were taken with both the 820-MHz and L-Band (1.4\,GHz) receivers with roughly monthly cadence. The dual-frequency observations are used to improve timing precision and dispersion measure (DM) estimates. Data collected prior to 2011 for PSR~\jten\ and PSR~\jtwentyone\ used the GASP pulsar backend (64\,MHz bandwidth with 4\,MHz frequency resolution) \citep{2007_GASP,NG15_timing}. All other data were collected using the Green Bank Ultimate Pulsar Processing Instrument (GUPPI) backend using 2048 frequency bins. For this backend the 820-MHz receiver has a 160\,MHz bandwidth with a time resolution of 81.92\,$\rm \mu$s and 2048 frequency bins. The L-Band receiver has a 640\,MHz bandwidth centered at 1500\,MHz with a time  resolution of 40.96\,$\mu$s and 2048 frequency bins. Times of arrival (TOAs) used in the timing analysis were published in the NANOGrav 15-yr data release \cite{NG15_timing}.

\subsection{CHIME} \label{subsec:chime_obs}

Data from the CHIME telescope used in this analysis were collected between April 2019 and March 2023 at nearly daily cadence. The CHIME/Pulsar backend collects 10 beamformed voltages with a frequency range of 400-800\,MHz, time resolution of 2.56\,$\rm \mu s$, and frequency resolution of 0.390625\,MHz \citep{chime_overview, chime_instrument}. Observation lengths were determined by a transit duration, typically 10--15 minutes for all sources. 

\subsection{Data Reduction} \label{subsec:data_reduction}

Data processing prior to the step of calculation of TOAs was done using the \verb+psrchive+\footnote{\url{https://psrchive.sourceforge.net/}} software suite with different procedure pipelines used for NG15 data vs CHIME data. For NG15 data this was done using the \verb+nanopipe+\footnote{\url{https://github.com/demorest/nanopipe}} data reduction pipeline which included radio frequency interference (RFI) excision, flux and polarization calibration, frequency scrunching to 64 channels, and time scrunching to a single sub-integration. 

For CHIME data, we developed a data processing pipeline \verb+chimera+ \citep{chimera} to perform equivalent processing steps to those in \verb+nanopipe+ specifically for generating WB TOAs. This pipeline is described in Appendix \ref{sec:chimera}. The exception being that we did not do polarization calibration of the CHIME data, as calibration methods for CHIME are still under development \citep{chime_instrument}. Polarization calibration would account for instrumental polarization distortions which could introduce systemic errors into TOA calculations if not corrected for \citep{van_straten_2026}. These errors can be particularly pronounced for sources with a high polarization fraction as the pulsar signal passing through the interstellar medium can also undergo significant Faraday rotation. Of the sources in this analysis, PSR~\jten\ and PSR~\jtwentyone\ have been previously observed to have significant polarization fractions of 0.66 and 0.21 respectively \citep{12.5yr_polarization}. Introducing polarization calibration in future analyses could account for instrumental errors, reduce TOA uncertainty, and improve timing rms residuals.

\subsection{WB Template Portraits} \label{subsec:wb_timing}

Generation of WB template portraits and TOA calculation was done using the \verb+PulsePoratraiture+\footnote{\url{https://github.com/pennucci/PulsePortraiture}} software suite \citep{pulse_portraiture}. WB template portraits are made by first constructing a high S/N data portrait by adding and aligning all observations from a given receiver together using \verb+ppalign.py+. From these data portraits, the template portraits are calculated using a combination of principal component analysis and spline interpolation using \verb+ppspline.py+. The choice of the number of significant eigenprofiles was done using an S/N threshold of 150, which was calculated using methods from \citet{pulse_portraiture} and  \citet{ng_nine_year_timing_og}. 

WB template portraits do not inherently account for instrument offsets and are not accurate beyond the observing frequency range of the data portrait \citep{pulse_portraiture}. The WB template portraits used for NG15 data were the same as those used in \citet{12.5yr_wideband}. We required the calculation of new CHIME-specific template portraits for each source. To avoid the inclusion of noisy or marginal detections, only observations with an S/N exceeding an empirically determined threshold, we chose the 50th percentile of S/N per pulsar, were included in the construction of the data portraits. This threshold was 60 for PSR~\jten\ and 20 for PSR~\jzerosix, \jtwentyone, and \jtwentythree.

\begin{table*}[]
% \centering
\caption{TOA Statistics and fit results. $N_{\rm toas}$ is the number of TOAs per dataset per pulsar.}
\begin{tabular*}{\textwidth}{@{\extracolsep{\fill}}lllll}
\toprule
% X*{9}{c}
Source                       & \jzerosix{}    & \jten{}    & \jtwentyone{}    & \jtwentythree{}    \\ \hline
CHIME $N_{\text{toas}}$         & 1099          & 1070          & 1036          & 1046          \\
NG15 $N_{\text{toas}}$          & 281           & 630           & 400           & 237           \\
MJD span                     & 55704 -- 60000 & 53267 -- 60000 & 53267 -- 60000 & 55972 -- 60000 \\
NG15 MJD span                & 55704 -- 58939 & 53267 -- 58939 & 53267 -- 58942 & 55972 -- 58942 \\
CHIME MJD span               & 58602 -- 60000 & 58700 -- 60000 & 58646 -- 60000 & 58646 -- 60000 \\
TOA rms ($\mu$s)        & 0.3097        & 0.6037        & 0.8337        & 1.9596        \\
DM rms (pc cm$^{-3}$) & 0.000155      & 0.000222      & 0.00109       & 0.000824      \\
$\chi^{2}_{\text{red}}$         & 1.858         & 1.924         & 2.300         & 1.098        \\ \hline
    % \endgroup 
\end{tabular*}
\label{tab:timing_stats}
\end{table*}

\begin{table*}[]
\centering
\caption{Timing Results for spin, binary, astrometric, and PK parameters. Reported DMs represent the central DM for the DMX model. DMX parameters are not listed for brevity. Derived parameters are indicated with an asterisk. $N_{\rm DMX}$ represents the number of DMX parameters used per pulsar for the preferred DMX model (Table~\ref{tab:dmx_binwidths}). Actual DMX values are omitted for brevity. }
\begin{tabular*}{\textwidth}{@{\extracolsep{\fill}}lllll}
\hline 
Source                                           & \jzerosix{}              & \jten{}               & \jtwentyone{}                  & \jtwentythree{}               \\ \hline
Right Ascension (J2000)                          & 6:45:59.082669(3)        & 10:12:33.438837(3)        & 21:45:50.457730(2)         & 23:02:46.9787911(3)                 \\
Declination  (J2000)                             & 51:58:14.88593(5)        & 53:07:02.18562(2)         & $-$7:50:18.52879(13)       & 44:42:22.06555(13)                   \\
$P$ (s)                                          & 0.008853496686733226(8) & 0.005255749022591625(8)  & 0.016052423674578594(3)   & 0.005192324649748711(12) \\
$\dot{P}\,(10^{-20}\, {\rm s ~ s}^{-1})$         & 0.4922052(9)             & 1.71274(2)                & 2.979272(5)                & 1.38704(3)               \\
DM (pc cm$^{-3}$)                                & 18.2476                  & 9.0224                    & 9.00369                    & 13.7234                  \\
$N_{\text{DMX}}$                                 & 404                      & 471                       & 438                        & 99           \\
$\mu_{\alpha}$ (${\rm mas ~ yr}^{-1}$)           & 1.516(11)                & 2.620(6)                  & $-$9.71(11)                & 0.0849(5)                 \\
$\mu_{\delta}$  (${\rm mas ~ yr}^{-1}$)          & $-$7.465(17)             & $-$25.529(4)              & $-$8.77(4)                 & $-$5.60(4)                   \\
$\varpi$ (mas)                                   & 0.75(9)                  & 1.0(10)                   & 1.58(5)                    & 0.6(6)                   \\ \hline
Binary Parameters                                &                          &                           &                            &                          \\ \hline
Binary Model                                     & \nodata                  & ELL1                      & ELL1H                      & ELL1H                    \\
$P_{\rm b} $(days)                               &  \nodata                 & 0.6046727138556(27)       & 6.838902509666(19)         & 125.935296927(6)         \\
$e$ *                                            &  \nodata                 &$1.46(8) \times 10^{-6}$   & $1.9320(6) \times 10^{-5}$ & 0.000503015(9)          \\
$\omega$ (deg) *                                 &  \nodata                 & 83.1(3.3)                 & 200.93(15)                 & 207.8855(12)            \\
$T_{\rm 0}$ (MJD) *                              &  \nodata                 & 56428.286(5)              & 56591.8268(4)              & 57813.8837(4)           \\
$x$ (ls)                                         &   \nodata                & 0.581817940(31)           & 10.16410970(4)             & 51.4299730(13)           \\
$\epsilon_{1}$                                   &  \nodata                 &$1.5(1) \times 10^{-6}$    & $-6.902(3) \times 10^{-6}$ & $-2.35248(4) \times 10^{-4}$\\
$\epsilon_{2}$                                   &   \nodata                &$8(111) \times 10^{-9}$    & $-1.8042(4) \times 10^{-5}$& $-4.44579(4) \times 10^{-4}$\\
$T_{\rm ASC}$ (MJD)                              &   \nodata                & 56339.86402951(12)        & 56588.009633654(3)         & 57993.02901376(5)        \\ \hline 
Post-Keplerian Parameters                        &                          &                           &                            &                          \\ \hline
$\dot{P}_{\rm b} \, (10^{-14})$                  &   \nodata                & 5.3(12)                   &    \nodata                 &     \nodata                      \\
$\dot{x} \,(10^{-15}\,{\rm ls~s}^{-1})$          &    \nodata               & 1.62(19)                  & 10.2(2)                    &     \nodata                      \\
$m_{\rm c}$ (${M}_{\odot}$) *                          &       \nodata            &   \nodata                 &     \nodata                & 0.32(9)                 \\
$\sin i$ *                                       &    \nodata               &   \nodata                 &   \nodata                  & 0.990(11)               \\
$h_3$ (s)                                        &   \nodata                &   \nodata                 & $6(1.3) \times 10^{-8}$    & $1.04(5) \times 10^{-6}$ \\
$\varsigma $                                       &   \nodata                &    \nodata                &     \nodata                & 0.84(2)  \\
NHARMS                                           &    \nodata               &     \nodata               & 3                          & \nodata                        \\ \hline
\end{tabular*}
\label{tab:timing_ephem}
\end{table*}

\section{Timing Analysis} \label{sec:timing_analysis}

Timing was done using the \verb+PINT+\footnote{\url{https://github.com/nanograv/PINT}} software package \citep{2021_PINT,2024_PINT}. To build the timing models for the combined CHIME/NG15 TOAs for each pulsar, we started with the NANOGrav 15-yr ephemeris and included new JUMP and DMJUMP parameters for the CHIME receiver. These account for any constant phase or DM offsets between receivers. Timing ephemerids are reported in Table~\ref{tab:timing_ephem} with fit results and TOA statistics reported in Table~\ref{tab:timing_stats}.

\subsection{Binary Modeling} \label{subsec:binary}

All three binary pulsars in this analysis PSR~\jten, PSR~\jtwentyone, and PSR~\jtwentythree, have nearly circular orbits. Periastron is poorly defined in orbits with low eccentricity, thus requiring the use of Lagrange Laplace terms $\epsilon_{1}$ and $\epsilon_{2}$ over the typical Keplerian parameters, eccentricity ($e$) and longitude of periastron ($\omega$) \citep{ell1_J1012}.

For both PSR~\jtwentyone\ and PSR~\jtwentythree\ we also have significant detections of Shapiro delay, requiring the use of the ELL1H model  \citep{ELL1H_model}, which uses an orthometric parameterization of higher order Shapiro delay harmonics. 
For PSR~\jtwentyone\ we were able to measure three Shapiro delay harmonics and thus used the medium inclination case which includes the the third harmonic ($h_{3}$) parameter 

\begin{equation} \label{eq:h3}
    h_{3} = r \left(\frac{s}{1+\sqrt{1-s^{2}}} \right)^{3}
\end{equation}

where $r$ and $s$ are the range and shape Shapiro delay parameters. We used the very high inclination parameterization for PSR~\jtwentythree\ which includes both the $h_{3}$ and the orthometric ratio ($\varsigma$) parameters:

\begin{equation} \label{eq:stigma}
     \varsigma =  \frac{s}{1+\sqrt{1-s^{2}}} 
\end{equation}

\subsection{DM Modeling} \label{subsec:dm_modeling}

Temporal DM variations were modeled using the piece-wise constant DMX model \citep{NG_nine_year_dm_Jones,NG15_timing}. In the wideband likelihood, the weighted mean of the wideband DM values in a DMX MJD range serves as the prior for fitting the DMX values \citep{12.5yr_wideband}. Thus, if there is significant variation in the number of observations per DMX range, this may artificially weight DMX values from denser ranges. The NG15 DMX models for the sources in this paper used 6.5-day DMX ranges for GUPPI TOAs and 15-day ranges for GASP TOAs, which were determined by the cadence of the dual frequency observations required for accurate DM measurements \citep{NG_nine_year_dm_Jones}. CHIME's wide 400\,MHz bandwidth is large enough to make independent DM measurements without the presence of dual frequency observations \citep{chime_instrument}. 

Similarly to \citet{Fonseca_J0740_2021}, we tested two categories of DMX ranges: constant and hybrid. In constant range models, we used the same DMX range width (6.5\,d, 15\,d, 30\,d) across the entire timing baseline. The only exceptions were for the two sources with GASP data (PSR~\jten\ and PSR~\jtwentyone) for which ranges including GASP TOAs had a minimum 15\,d length. As DMX models add a large number of parameters to the timing model, the longer DMX ranges were used to test if we may be able to sufficiently model slowly evolving DM variations with fewer parameters.

In hybrid range models we used combinations of longer DMX ranges for data prior to MJD 58600 (30\,d, 15\,d, 6.5\,d) and shorter ranges after (15\,d, 6.5\,d, 3\,d) to account for the higher cadence of CHIME data (Table~\ref{tab:dmx_binwidths}). Model selection was done using the Aikaike Information Criterion (AIC) statistic \citep{BurnhamAnderson2004_AIC}. The DMX model with the lowest AIC for each source was the ``preferred" model and other models were discarded if the relative likelihood ($\exp (-\Delta \text{AIC}/2)$) was below 0.05 which corresponds to a $\Delta \text{AIC}$ of roughly 6. For PSR~\jzerosix\ and PSR~\jten\ the preferred DMX model was a hybrid range model with 6.5\,d/3\,d ranges pre/post MJD 58600. The preferred model for PSR~\jtwentythree\ was a constant range model with 30\,d bin widths. Timing results reported in Table~\ref{tab:timing_stats} are using the preferred DMX model respective to each pulsar. 

\begin{figure*}
    \centering
    \includegraphics[width=\textwidth]{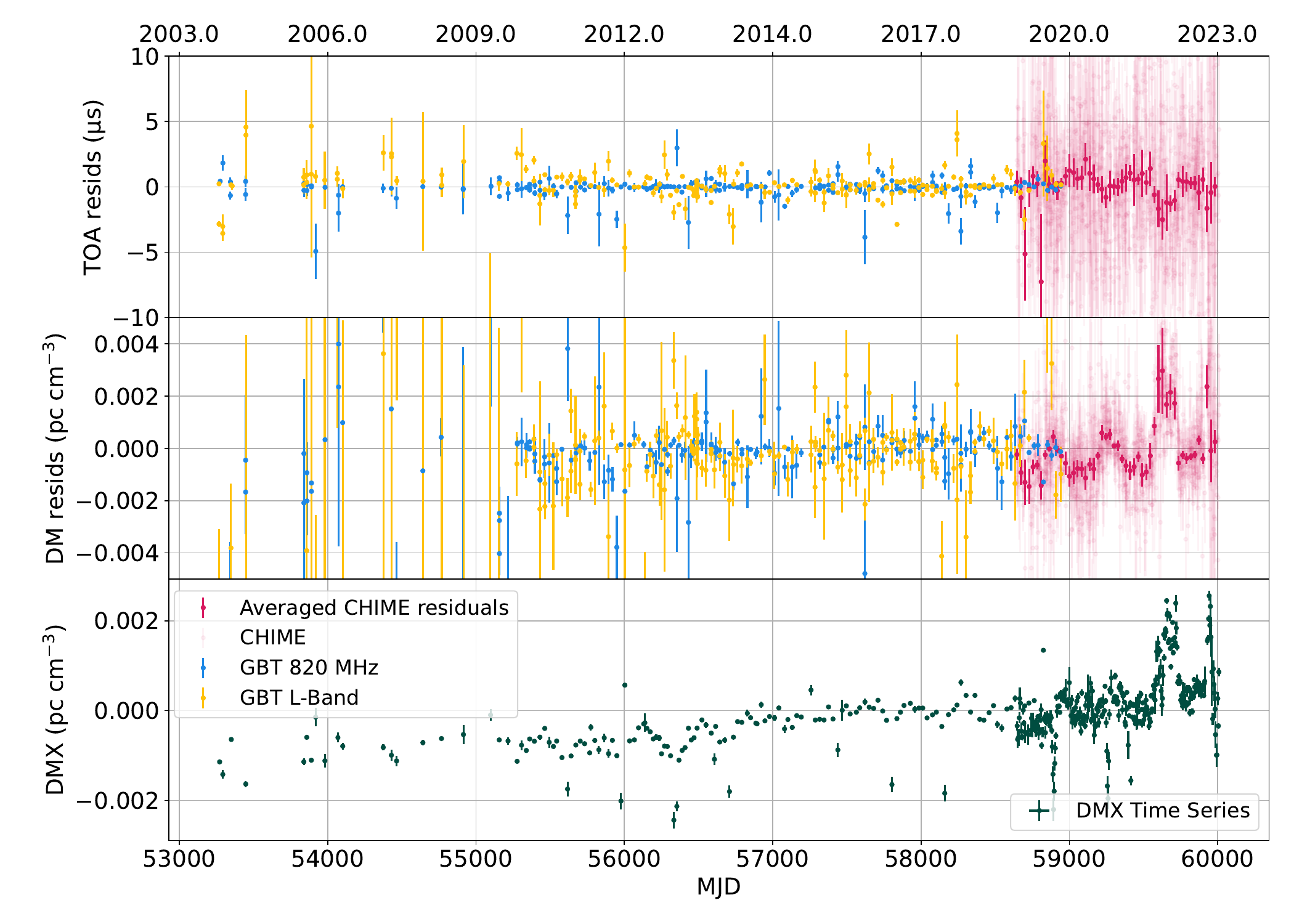}
    \caption{Top plot: timing resids for PSR~\jtwentyone; Middle plot: DM residuals; Bottom plot: DMX time series where vertical bars indicate DMX parameter uncertainty and horizontal bars indicate the width of DMX bins in time.  }
    \label{fig:J2145_resids}
\end{figure*}

PSR~\jtwentyone\ has two DMX models from which we cannot distinguish a preferred model with the AIC statistic alone. The hybrid 6.5\,d/3\,d range model technically has the lowest AIC value, but there is only a $\Delta \text{AIC}$
 of 1 with the 15\,d/3\,d range model. This gives us a relative likelihood of 0.6 for the 15\,d/3\,d model to minimize information loss compared to the the 6.5\,d/3\,d model. Even though the maximum DMX range for pre MJD 58600 data is more than doubled in the 15\,d/3\,d model, the roughly monthly cadence of observations in this period means there are only seven fewer DMX parameters than the 6.5\,d/3\,d model. Given the very large number of total DMX parameters in each model (445 vs.\ 438) these models are not different enough from each other for one to be preferred over the other. The more extreme temporal DM variations for this source are in post MJD 58600 data which has the same DMX range width for both models (Fig.~\ref{fig:J2145_resids}). Since neither is preferred we decided to select the 15\,d/3\,d model as it has slightly fewer free parameters. 

\begin{table*} \label{tab:aic_results}
    \centering
    \caption{$\Delta$AIC values for each set of DMX model parameters, where $\Delta \text{AIC} = \text{AIC} - \text{AIC}_{\min}$. A $\Delta$AIC value of zero indicates the DMX model with the lowest AIC value for each source.}
    \begin{tabular*}{\textwidth}{@{\extracolsep{\fill}}lllll}
    \hline
        Model                    & \jzerosix{}  & \jten{}  & \jtwentyone{}  & \jtwentythree{}  \\ \hline
        6.5 day                  & 36   & 85   & 856   & 144 \\
        15 day                   & 71   & 266  & 2659  & 65 \\
        30 day                   & 1455 & 1266 & 10649 & 0 \\
        6.5 day NG15 3 day CHIME & 0    & 0    & 0     & 217    \\ 
        15 day NG15 3 day CHIME  & 9    & 29   & 1     & 218    \\
        15 day NG15 6.5 day CHIME& 45   & 105  & 755   & 144   \\
        30 day NG15 3 day CHIME  & 97   & 878  & 2295  & 190  \\
        30 day NG15 6.5 day CHIME& 138  & 1007 & 3108  & 113 \\
        30 day NG15 15 day CHIME & 163  & 1168 & 5409  & 36      \\ \hline
    \end{tabular*}
    \label{tab:dmx_binwidths}
\end{table*}

\subsection{Noise Modeling} \label{subsec: noise}

Analysis of noise properties in the TOA and DM residuals was done using the same methodology as \citet{12.5yr_wideband} and \citet{NG15_timing} using the \verb+enterprise+ Bayesian pulsar timing suite \citep{enterprise_software}. Once both white and red noise parameters are determined with \verb+enterprise+, they are included in the timing ephemeris as frozen parameters. 
For each receiver/backend combination, white noise was parameterized using EFAC, DMEFAC, EQUAD, and DMEQUAD. EFAC is a linear scaling factor for TOA uncertainties used to account for mismatches between the template portrait and the data, and EQUAD accounts for additive white noise and is added in quadrature to TOA uncertainties. DMEFAC and DMEQUAD are analogous to EFAC and EQUAD, but applied to the WB TOA DM measurements \citep{12.5yr_wideband}. Red noise is modeled as a stationary Gaussian process with power spectrum:
\begin{equation}
    P(f) = A^{2}_{\rm red} \left(\frac{f}{\rm 1yr^{-1}}\right)^{\gamma_{\rm red}}
\end{equation}

Where $A_{\rm red}$ is the power spectrum amplitude at a frequency of 1 yr$^{-1}$\textit{, f is the Fourier frequency,} and $\gamma_{\rm red}$ is the spectral index.  Similarly to \citet{12.5yr_wideband}
we considered detection of red noise to be significant if the Savage-Dickey density ratio of the posterior distribution of $\log_{10}(A_{\rm red})$ exceeded 100 \citep{Dickey1971}. We detected significant red noise in PSR~\jten, which was also detected in NG15. We get somewhat different values for $\gamma_{\rm red}$, but the NG15 value of $-0.8(3)$ is consistent within the 2-$\sigma$ confidence interval of the CHIME/NG15 result of $-1.3(4)$. The Fourier frequencies used to fit for $A_{\rm red}$ and $\gamma_{\rm red}$ are dependent on the total timing baseline, which is longer by nearly three years for CHIME/NG15, thus we fitting these parameters using somewhat different frequencies. Additionally, the added CHIME data has both a higher observing cadence and increased TOA uncertainties, which implicitly changes the weighting of the underlying basis functions used for fitting. Both PSR~\jzerosix\ and PSR~\jtwentythree\ did not have significant detections of red noise in both this analysis and NG15.
For PSR~\jtwentyone, there was significant red noise in the NG15 data, but with the combined CHIME/NG15 dataset our attempts to model red noise would introduce large, non-physical sinusoids in our timing residuals so we excluded it from the noise model and only used white noise parameters. We believe this difficultly may be due to the strong solar wind variability we see the DM times series (Sections~\ref{subsec:dm_modeling} \&~\ref{subsec:swx}). If we had been able to model red noise we would expect to see some increases in timing parameter uncertainty and possibly some changes in post-keplerian parameter measurements as unmodeled red noise processes would vary on similar timescales to relativistic post-keplerian processes. For PSR~\jten\ we did observe some changes to post-keplerian parameters orbital decay and rate of change of the projected semi-major axis prior to and after introducing a red noise model that were within 2-$\sigma$ measurement uncertainties.

\section{Results} \label{sec:results}

\subsection{PSR~\jzerosix\ }

We report marginal improvement in precision of the updated timing astrometric measurements for the isolated PSR~\jzerosix. Our new parallax distance of 1.2(2) kpc which has 2-$\sigma$ agreement with the NG15 distance of 1.5(3)\,kpc. Using our proper motion measurements we determined a total proper motion of $7.62 \pm 0.02 \text{\,mas yr}^{-1}$ and a corresponding transverse velocity of $45 \pm 8 \text{\,km s}^{-1}$.

Our $\dot{P}$ measurement was mostly due to intrinsic pulsar spin down ($\dot{P}^{\text{Int}}$) but we did estimate significant extrinsic contributors: 
 \begin{equation} \label{eq:pdot_tot}
     \dot{P} = \dot{P}^{\text{Int}} + \dot{P}^{\text{Shk}} + \dot{P}^{\text{Gal,rad}} + \dot{P}^{\text{Gal,az}}
 \end{equation}

This includes the kinematic term contribution due to transverse velocity of the pulsar ($\dot{P}^{\text{Shk}}$) and  acceleration due to differential galactic rotation ($\dot{P}^{\text{Gal,rad}}$), and acceleration towards the galactic disk ($\dot{P}^{\text{Gal,az}}$) \citep{Shklovskii_orig,Damour_Taylor_1991,Nice_Taylor_1995}. 

We estimated $\dot{P}^{\text{Shk}}$ to be $2.1(4) \times 10^{-22} \text{\,s\,s}^{-1}$ using:

\begin{equation}\label{eq:shklovskii_term}
    \dot{P}^{\text{Shk}} = P \frac{(\mu_{\alpha}^{2} + \mu_{\delta}^{2})D}{c}
\end{equation}

Where $\mu_{\alpha} \equiv \frac{d \alpha}{dt} \cos\delta$ is the proper motion in right ascension, $\mu_{\delta}$ is the proper motion in declination, and $D$ is the parallax distance.

We estimated $\dot{P}^{\text{Gal, rad}}$ to be $6.6(1.1) \times 10^{-22}\text{\,s\,s}^{-1}$ using:

\begin{equation} \label{eq:pdot_gal_rad}
    \dot{P}^{\text{Gal, rad}} = P \left[ \frac{-\Omega_{0}}{c R_{0}} \cos b \left( \cos l + \frac{\beta}{\beta^{2}+\sin^{2}l}   \right) \right]
\end{equation}

Where $l$ and $b$ are the Galactic coordinates of the pulsar, $d$ is the parallactic distance, and $\beta \equiv\frac{d}{R_{0}}\cos b - \cos l$. We used a Sun-Galactic Center distance $R_0 = 8.275 \pm 0.033$\,kpc \citep{Gravity_2021} and a circular speed of the local standard of rest of $\Omega_{0} = 233.3 \pm 1.4$\,km\,s$^{-1}$ \citep{McGaugh_2018}. 

We estimated $\dot{P}^{\text{Gal,az}}$ to be $-4.1(4) \times 10^{-22} \text{\,s\,s}^{-1}$ using:

\begin{equation}
    \dot{P}^{\text{Gal,az}} = P \left[ \frac{K_{z} |\sin b|}{c}  \right]
\end{equation}

Where $K_{z}$ is the vertical component of Galactic acceleration. We have used the approximation for $K_{z}$ in Equation 17 in \citet{Lazaridis_wex_2009} which is accurate for Galactic heights $z \equiv |d \sin b| \leq 1.5 \text{\,kpc}$ \citep{2004_Holmberg_Flynn}. This source has a Galactic height of roughly 0.43\,kpc. 

This leaves us with a remaining intrinsic $\dot{P}$ of $4.5(12) \times 10^{-21} \text{ s s}^{-1}$.

\subsection{PSR~\jten\ } \label{subsec:J1012_results}

Measurement of orbital decay ($\dot{P}_{B}$) for PSR~\jten\ improved in precision by a factor of five compared to the NG15 measurement of $6.1(5) \times 10^{-14} \text{ s s}^{-1}$ \citep{NG15_timing}. We looked at several potential intrinsic and extrinsic contributors to the observed total $\dot{P}_{B}$:

\begin{equation}\label{eq:pbdot_tot}
     \dot{P}_{B}^{\rm Tot} = \dot{P}_{B}^{\rm GR} + \dot{P}_{B}^{\rm Shk} + \dot{P}_{B}^{\rm Gal} + \dot{P}_{B}^{\dot{m}} + \dot{P}_{B}^{\rm T} + \dot{P}_{B}^{\rm ex} 
\end{equation}

The emission of gravitational waves ($\dot{P}_{B}^{\rm GR}$), mass loss from the system ($\dot{P}_{B}^{\dot{m}}$), and tidal dissipation of the orbit ($ \dot{P}_{B}^{\rm T}$) are all intrinsic to the binary. Extrinsic contributors include transverse acceleration of the binary system ($\dot{P}_{B}^{\rm Shk}$), and acceleration due to the Galactic gravitational potential ($\dot{P}_{B}^{\rm Gal}$) \citep{Damour_Taylor_1991,Nice_Taylor_1995,Shklovskii_orig}. For this system $\dot{P}_{B}^{\dot{m}}$ and $\dot{P}_{B}^{\rm T}$ are negligible.

The most significant contributor to $\dot{P}_{B}^{\rm Tot}$ measurement for PSR~\jten\ was the Shklovskii term which we estimated using Equation~\ref{eq:shklovskii_term} with $ \dot{P}_{B}$ and $P_B$ instead of $\dot{P}$ and $P$ \citep{Shklovskii_orig}. 

The uncertainty of $\dot{P}_{B}^{\rm Shk}$ is dominated by the timing parallax measurement which, for PSR~\jten, was consistent within 1.4-$\sigma$ to Very Long Baseline Interferometry (VLBI) determined parallax measurements from \citet{2023_ding_deller} (Table~\ref{tab:astrometry_J1012}). The larger uncertainty in the CHIME/NG15 timing parallax vs VLBI parallax explains the larger uncertainty of our estimate of $\dot{P}_{B}^{\rm Shk}$, though still consistent within 1.1-$\sigma$ of the $\dot{P}_{B}^{\rm Shk}$ estimate from VLBI astrometry (Table~\ref{tab:pbdot_J1012}).

We estimated $ \dot{P}_{B}^{\rm Gal}$ using methods from \citet{Nice_Taylor_1995} and \citet{Lazaridis_wex_2009}. 

\begin{equation}
    \dot{P}_{B}^{\rm Gal} = \frac{A_{G}}{c}P_{B}
\end{equation}

$A_{G}$ is the line of sight acceleration of the pulsar binary which we calculated with the same values of $R_0$ and $\Omega_{0}$ used in Equation~\ref{eq:pdot_gal_rad}.

\begin{table}[] 
\centering
\caption{Astrometry comparison for \jten}
\begin{tabularx}{\columnwidth}{llll}
\hline
Analyses                    & $\mu_{\alpha}$ (mas yr$^{-1}$) & $\mu_{\delta}$ (mas yr$^{-1}$)& $\varpi$ (mas)  \\ \hline
This work                   & 2.620(6)                       & -25.529(6)                    & 1.0(10)            \\
\citet{J1012_vlbi}          & 2.67(9)                        & -25.40(14)                    & 1.21(8)          \\
\citet{2023_ding_deller}    & 2.61(1)                        & -25.49(1)                     & 1.14(4)          \\
 \hline
\end{tabularx}
\label{tab:astrometry_J1012}
\end{table}

\begin{table*}[] 
\centering
\caption{$\dot{P}_{B}$ contributor comparison for \jten. Single asterisk indicates measurement carried over from \citet{J1012_vlbi}. Double asterick indicates measurement carried over from \citet{2023_ding_deller}. As VLBI distances are more precise and reliable than those determined by pulsar timing we also calculated contributions to our $\dot{P}_{B}^{\rm Tot}$ measurement using VLBI-determined parallax and proper motion terms.}
\begin{tabular*}{\textwidth}{@{\extracolsep{\fill}}llllll}
\hline
Analyses                    & $\dot{P}_{B}^{\rm Tot}$ (fs s$^{-1}$) & $\dot{P}_{B}^{\rm Shk}$ (fs s$^{-1}$) & $\dot{P}_{B}^{\rm Gal}$ (fs s$^{-1}$)& $\dot{P}_{B}^{\rm GR}$ (fs s$^{-1}$) & $\dot{P}_{B}^{\rm ex}$ (fs s$^{-1}$)\\ \hline
This work                   & 53(1)                                 & 82(8)                                 & -5.8(3)                             &   -23(8)                                  &                                     \\
This work - VLBI corrections  & 53(1)                                 & 73(3)**                                 & -4.1(4)**                             &   -15(3)                                  &                                     \\
\citet{J1012_vlbi}          & 61(4)$^{a}$                                 & 68.6(4.4)                             & -5.5(2)                             & -13(1)                               &     10.6(6.1)                                \\
\citet{2023_ding_deller}    & 61(4)$^{a}$                                & 73(3)                                 & -4.1(4)                              & -13(1)*                              &       -7.9(5.0)                              \\
 \hline
\end{tabular*}
\small
\begin{center}
\begin{tabular}{@{}l@{}}

$^a$\citet{2016_desvignes}
\end{tabular}
\end{center}
% \item $^a$\citet{2016_desvignes}
\label{tab:pbdot_J1012}
\end{table*}

This gave us a remaining excess $\dot{P}_{B}$ term of $-23(8)\,\text{fs\,s}^{-1} $. While this excess term is only marginally significant due to the high uncertainty, we hypothesized that it was likely due to $ \dot{P}_{B}^{\rm GR}$. We re-did our calculation of this excess term using the $ \dot{P}_{B}^{\rm Shk}$ and $ \dot{P}_{B}^{\rm Gal}$ terms calculated from VLBI astrometry and this time got a much more precise value of $-15(3)\,\text{fs\,s}^{-1}$. This term is consistent with the estimated $\dot{P}_{B}^{\rm GR}$ from \citet{J1012_vlbi} which itself was made using the pulsar-companion mass ratio reported in \citet{J1012_binary_mass} and Equation 21 in \citet{Lazaridis_wex_2009}.

\begin{figure}
    \centering
    \includegraphics[width=\columnwidth]{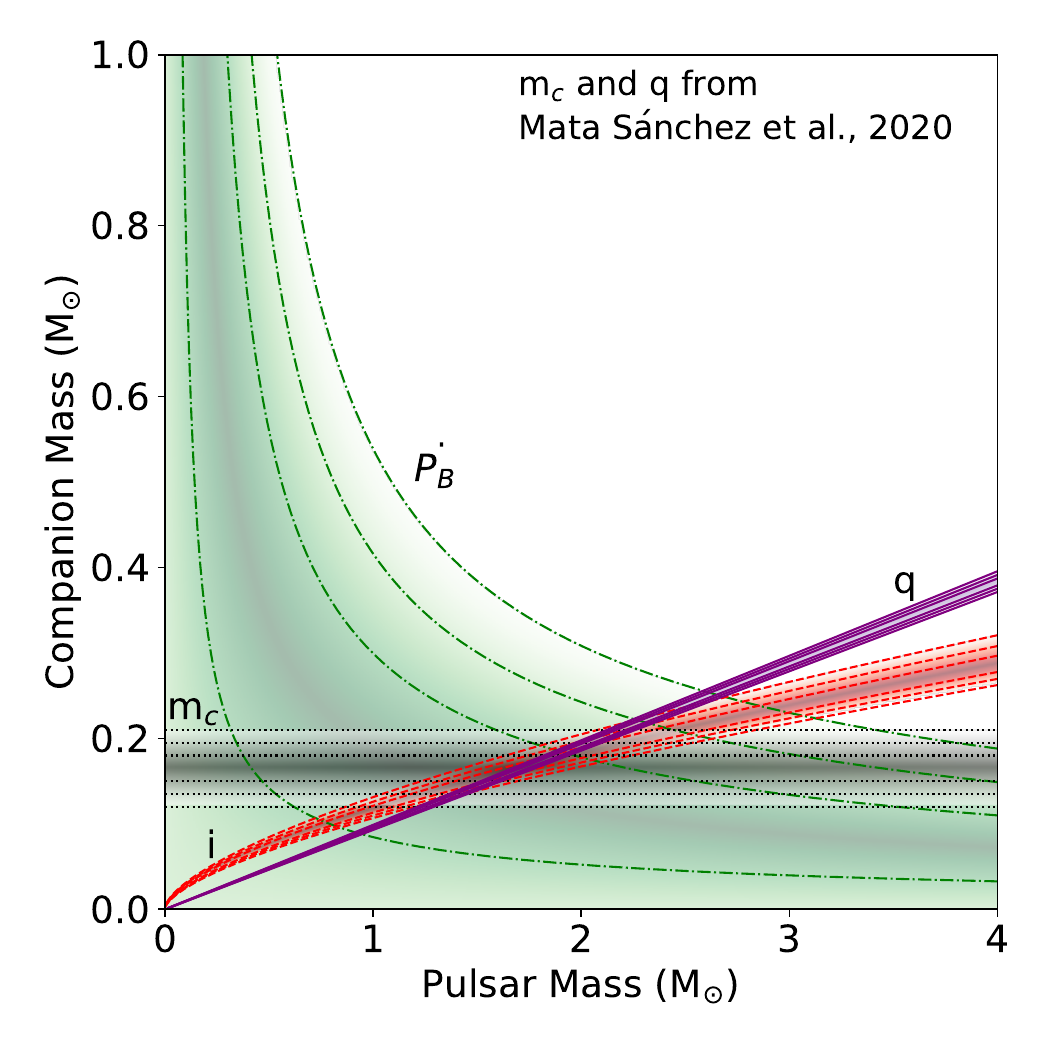}
    \caption{Pulsar and companion mass constraints for PSR~\jten. Green contours are from the estimated $\dot{P_B}^{\rm GR}$ term and red contours are from inclination angle constraint determined from the kinematic term of $\dot{x}$. Both the companion mass ($m_{\rm c}$) and pulsar-companion mass ratio ($q$) contours are determined from measurements from \citet{J1012_binary_mass}}
    \label{fig:J1012_mass_mass}
\end{figure}

We then treated the VLBI corrected excess $\dot{P}_{B}$ term as a $\dot{P}_{B}^{\rm GR}$ estimate and looked at the implied curve in pulsar and companion mass space (Fig.~\ref{fig:J1012_mass_mass}). The 1-$\sigma$ region directly overlaps with intersection of the companion mass and pulsar-companion mass ratio curves calculated from measurements reported in \citet{J1012_binary_mass}.

The measurement of rate of change of the projected semi-major axis ($\dot{x}$) for PSR~\jten\ improved in precision by a factor of three from the NG15 value. Using our own measurement uncertainty, the NG15 value is 5 standard deviations from our measurement, suggesting the values may be significantly different, but when using the larger NG15 measurement uncertainty, our measurement was within two standard deviations. Thus we can treat these measurements as broadly consistent, with the CHIME/NG15 measurement having improved precision due to the longer timing baseline. Contributors to the measured value of $\dot{x}$ are summarized below:

\begin{equation} \label{eq:xdot_tot}
    \dot{x}^{\rm Tot} = \dot{x}^{\rm GR} + \dot{x}^{\rm PM} + \frac{\text{d}\epsilon_{\rm A}}{\text{d}t} - \frac{\dot{D}}{D}+ \dot{x}^{\dot{m}} + \dot{x}^{\rm SO} + \dot{x}^{\rm planet}
\end{equation} 

Here $\dot{x}^{\rm GR}$ is contribution due to the emission of gravitational waves, $\dot{x}^{\rm PM}$ is from proper motion of the binary system, $\frac{\text{d}\epsilon_{\rm A}}{\text{d}t}$ is varying aberration due to geodetic precession of the pulsar spin axis, and $\frac{\dot{D}}{D}$ is changing Doppler shift. $\dot{x}^{\dot{m}}$ is from mass loss of the system, $\dot{x}^{\rm SO}$ is from spin-orbit coupling, and $\dot{x}^{\rm planet}$ in cases of a third companion. 

\begin{figure}
    \centering
    \includegraphics[width=\linewidth]{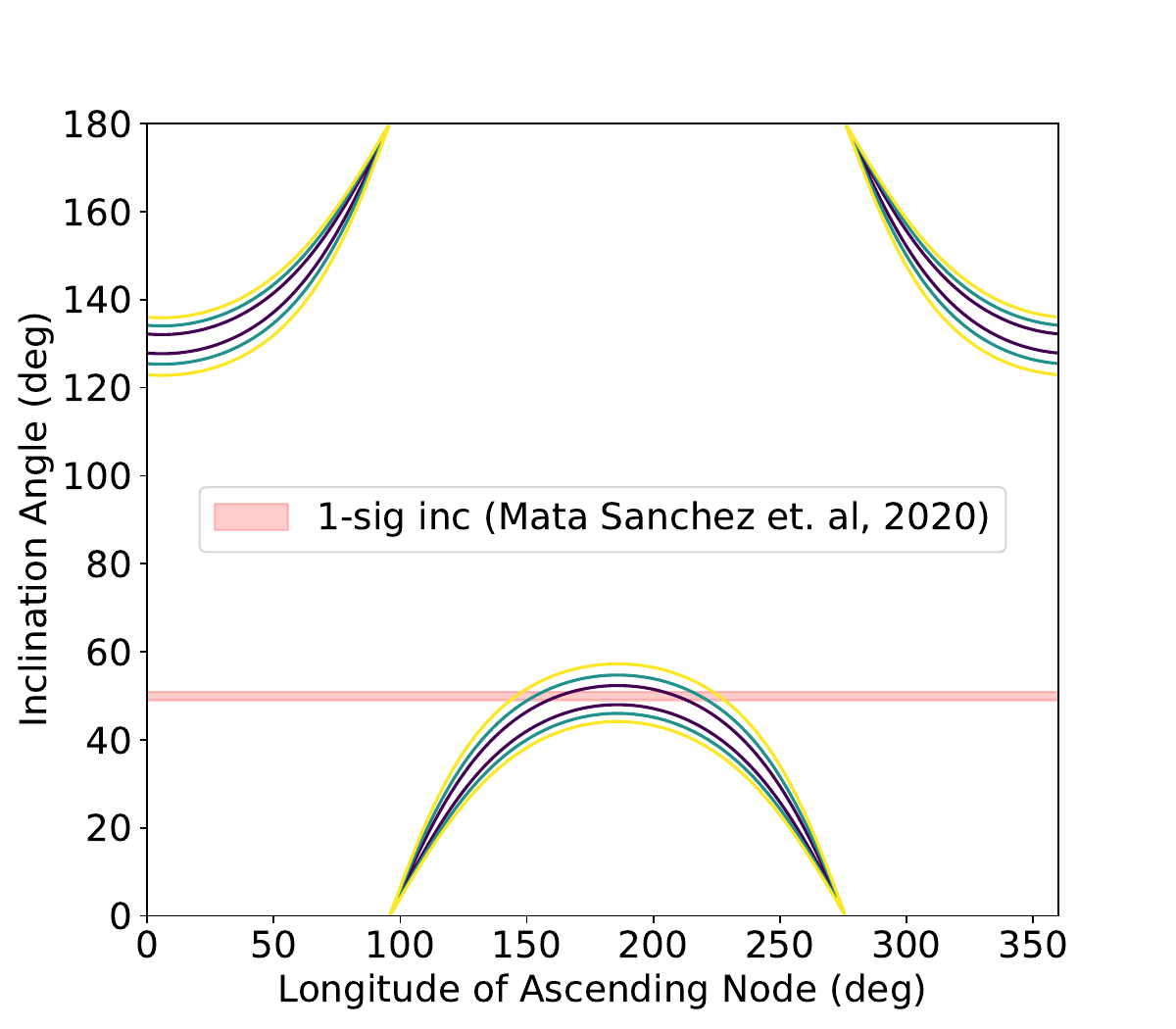}
    \caption{Contours of $\Omega_{\text{asc}}$ and $i$ from $\dot{x}$ measurement for \jten}
    \label{fig:J1012_om_asc}
\end{figure}

Our $\dot{x}$ measurement was dominated by $\dot{x}^{\rm PM}$. As all other estimated contributions are significantly smaller than the measurement uncertainty, we treated our timing measurement of $\dot{x}$ as an upper limit for the $\dot{x}^{\rm PM}$ contribution. We used this to look at the combinations of the longitude of ascending node ($\Omega_{\text{asc}}$) and the orbital inclination ($i$) implied by this $\dot{x}^{\rm PM}$ estimate and found they are partially consistent with the $i = 50\degree \pm 2\degree$ measurement reported in \citet{J1012_binary_mass}. When making no assumptions about which quadrant of $i$ and $\Omega_{\text{asc}}$ is correct, we found that in the $90-180\degree$ quadrant, $i$ must be above $120\degree$ (Fig.~\ref{fig:J1012_om_asc}).

We calculated more direct constraints on $i$ in the $0-90\degree$ quadrant from the following limit \citep{pulsar_handbook,Kopeikin_1996}: 

\begin{equation} \label{eq:inc_limit}
    \tan i  \leq 1.54 \times 10^{-16} \left( \frac{\mu^{\rm T}}{\rm mas\ yr^{-1}} \right) \left( \frac{x}{\dot{x}^{\rm PM}} \right)_{\rm max}
\end{equation}

From this we calculated the upper limit $\tan i \leq1.19(9)$. In the $0-90\degree$ quadrant, this gave a limit of $i \leq 52\degree \pm  4\degree$. Although we treated this estimate as a limit and not a measurement of $i$, as this method of estimation does not constrain the longitude of ascending node, we noted that it is still consistent within 1-$\sigma$ of the \citet{J1012_binary_mass} $i$ measurement. We used the binary mass function to plot the combinations of pulsar and companion masses implied by our $i$ limit in Figure~\ref{fig:J1012_mass_mass}. We showed intersection with the contours for $q$, $m_{\rm c}$, and $\dot{P}_{B}$ over a pulsar mass range of 1.3--2.1\,$M_{\odot}$ which is broadly consistent with the literature measurement of $1.72\pm 0.16 \,M_{\odot}$ \citep{J1012_binary_mass}. 

\subsection{PSR~\jtwentyone\ } \label{subsec:J2145_results}

\begin{figure}
    \centering
    \includegraphics[width=\columnwidth]{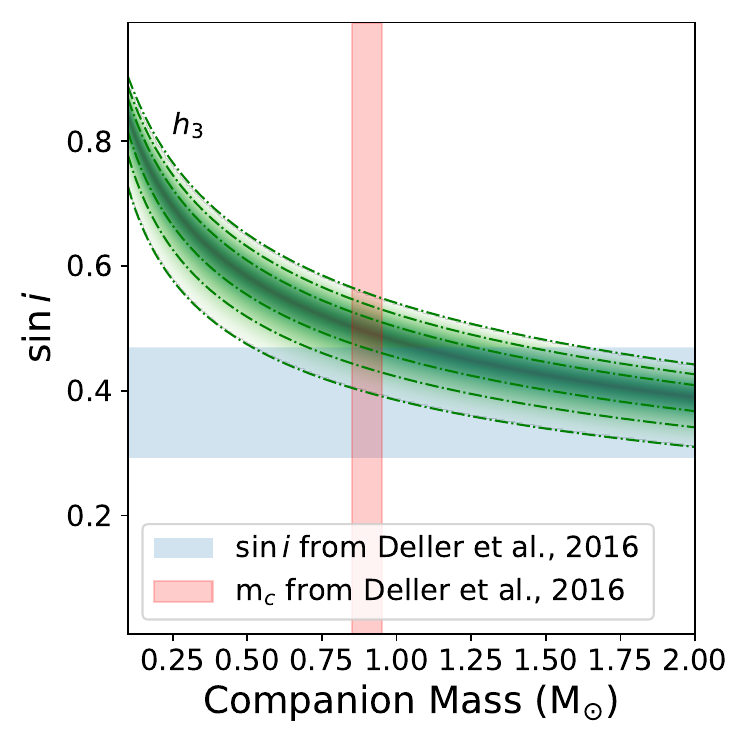}
    \caption{In green, the curve of $m_{\rm c}$ vs $\sin i$ from detected value of $h_3$ for PSR~\jtwentyone. The dashed lines represent the 1, 2, and 3-$\sigma$ error regions. Red  and blue shaded region represent the 1-$\sigma$ range of $m_{c}$ and $\sin i$ measurements respectively from \citet{J2145_mass_measurement}}
    \label{fig:J2145_shap_delay}
\end{figure}

Our detection of the $h_3$ Shapiro delay parameter is not sufficient to directly determine $m_{\rm c}$ and $\sin i $, as we could not detect any higher harmonics. Instead we defined a curve in $m_{\rm c}$ and $\sin i $ space implied by our $h_{3}$ measurement using Eq.~\ref{eq:h3} (Fig.~\ref{fig:J2145_shap_delay}). We compared to results from \citet{J2145_mass_measurement}, which used a combination of VLBI observations and optical photometry of the WD companion to determine companion mass, inclination, and distance. As shown in Figure~\ref{fig:J2145_shap_delay} the intersection in $m_{c} - \sin i$ space is consistent within the 1-$\sigma$ error region of our $h_3$ measurement.

\begin{figure}
    \centering
    \includegraphics[width=\linewidth]{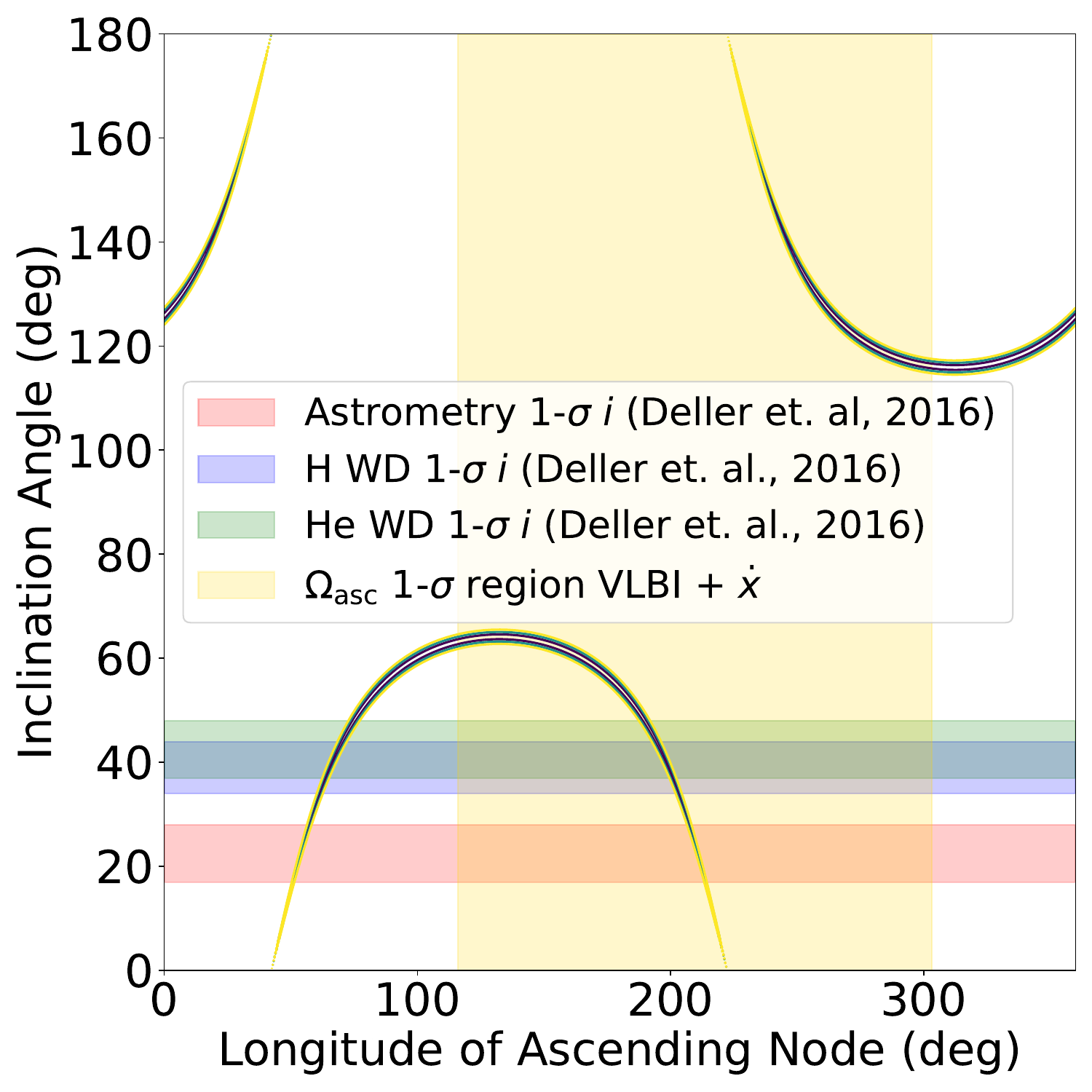}
    \caption{Contours of $\Omega_{\text{asc}}$ and $i$ calculated from $\dot{x}$ measurement of PSR~\jtwentyone. The gold region represent the 1-$\sigma$ error region of $\Omega_{\text{asc}}$ determined by calculation of reflex motion of VLBI data using our $\dot{x}$ measurement as a prior. }
    \label{fig:J2145_om_asc}
\end{figure}

The CHIME/NG15 $\dot{x}$ measurement improved in precision by a factor of 4 from the NG15 measurement, but has changed in value and is 12.5-$\sigma$ from the NG15 value of \citep{NG15_timing}. Looking at other analyses of this pulsar, such as \citet{ppta_timing} and \citet{ipta_dr2_timing}, our $\dot{x}$ is more precise but has significantly increased (Table~\ref{tab:astrometry_J2145}). Our analysis uses a much longer timing baseline than both \citet{NG15_timing} and \citet{ppta_timing}, which may explain the improved precision from these two analyses. It is possible that since we were unable to model red noise, which has been significantly detected in this source in previous analyses and can impact post-keplerian parameters if unmodeled, this maybe the reason we detect a significantly higher $ \bf \dot{x}$ value.
Similarly to PSR~\jten, the $\dot{x}$ measurement for PSR~\jtwentyone\ is dominated by the $\dot{x}^{\text{PM}}$ term and all other estimated contributors are smaller than the measurement uncertainty. Using the same methods as in Section~\ref{subsec:J1012_results}, we show that the implied combinations of $i$ and $\Omega_{\text{asc}}$ broadly overlap within the 0$\degree$--90$\degree$ $i$ quadrant with VLBI determined measurements from \citet{J2145_mass_measurement} of $i$ (Fig.~\ref{fig:J2145_om_asc}). Again using Equation~\ref{eq:inc_limit} we calculated an upper limit of $\tan i \leq 1.98(4)$ which corresponds to an $i$ limit of $63(4)\degree$ in the 0$\degree$--90$\degree$ $i$ quadrant.

\begin{table*}[] 
\centering
\caption{Astrometry comparison for PSR~\jtwentyone, with last digit uncertainties in parentheses. Red text indicates values reported without measurement uncertainties. }
\begin{tabular*}{\textwidth}{@{\extracolsep{\fill}}lllll}
\hline
Analyses                    & $\mu_{\alpha}$ (mas yr$^{-1}$) & $\mu_{\delta}$ (mas yr$^{-1}$)& $\varpi$ (mas) & $\dot{x}$ ($10^{-15} \ \rm{ls~s}^{-1}$) \\ \hline
This work                   & $-$9.71(11)                              & $-$8.54(4)                             & 1.55(5)            &   10.2(2)     \\
\citet{NG15_timing}         & $-$9.54(4)                              & $-$8.9(17)                              & 1.5(18)           & 7.5(8)       \\
 \citet{ppta_timing}        & $-$9.48(2)                              & $-$9.11(7)                              & 1.40(8)           & 6.1(4)       \\
\citet{J2145_mass_measurement}& $-$9.46(5)                            & $-$9.08(6)                              & 1.63(4)           &  \nodata         \\
\citet{ipta_dr2_timing}    & $-$9.58(3)                               & $-$8.87(7)                              & 1.54(10)          & 7.5(5)       \\
\citet{mpta_timing}        & \textcolor{red}{$-$9.46}                 & \textcolor{red}{$-$8.08}                & \textcolor{red}{1.33} & $-$35(5)   \\
\citet{mpta_astrometry}    & $-$9.7(3)                                & $-$8.2(10)                              & 1.6(3)            &  \nodata            \\
 \hline
\end{tabular*}
\label{tab:astrometry_J2145}
\end{table*}

\subsection{PSR~~\jtwentythree\ } \label{subseb:J2302_results}

PSR~\jtwentythree\ has a significant detection of Shapiro delay from which we determined the companion and pulsar masses, as well as the system inclination. To determine robust credible intervals for $m_{\rm c}$, $m_{\rm p}$, and $i$, we used the $\chi^2$ gridding method described in \citet{Splaver_2002}. We started with a range of values from uniform distributions of $m_{\rm c}$ and $\cos i$ and then calculated corresponding values of $h_3$ and $\varsigma$ using equations \ref{eq:h3} and \ref{eq:stigma}. These $h_3$ and $\varsigma$ pairs are then inserted in the timing model as frozen parameters and the timing model is re-fit with all other free parameters allowed to vary and the $\chi^2$ recorded. This $\chi^2$ grid is then mapped into the following likelihood function where $\chi^{2}_{0}$ is the minimum $\chi^2$: 

\begin{equation}\label{eq:h3stigma_pdf}
    p(h_3, \varsigma | \text{data}) = \exp\left(-\frac{(\chi^2 - \chi^{2}_{0})}{2} \right)
\end{equation}

Using Bayes's theorem and equations \ref{eq:h3} and \ref{eq:stigma} we can then transform this likelihood into a posterior PDF for $m_{\rm c}$ and $\cos i$:

\begin{equation} \label{eq:mccosi_pdf}
    p(m_{\rm c} , \cos i | \text{data}) = \left|\frac{\partial h_3}{\partial m_{\rm c}} \frac{\partial \varsigma}{\partial \cos i}\right| p(h_3,\varsigma | \text{data})
\end{equation}

We can then use the Keplerian mass function to translate this into a 2D-PDF for $m_{\rm p}$ and $\cos i $.

\begin{equation} \label{eq:mpcosi_pdf}
    p(m_{\rm p},\cos i | \text{data}) = \left|\frac{\partial m_{\rm c}}{\partial m_{\rm p}}\right|p(m_{\rm c}, \cos i | \text{data}) 
\end{equation}

\begin{figure}
    \centering
    \includegraphics[width=\linewidth]{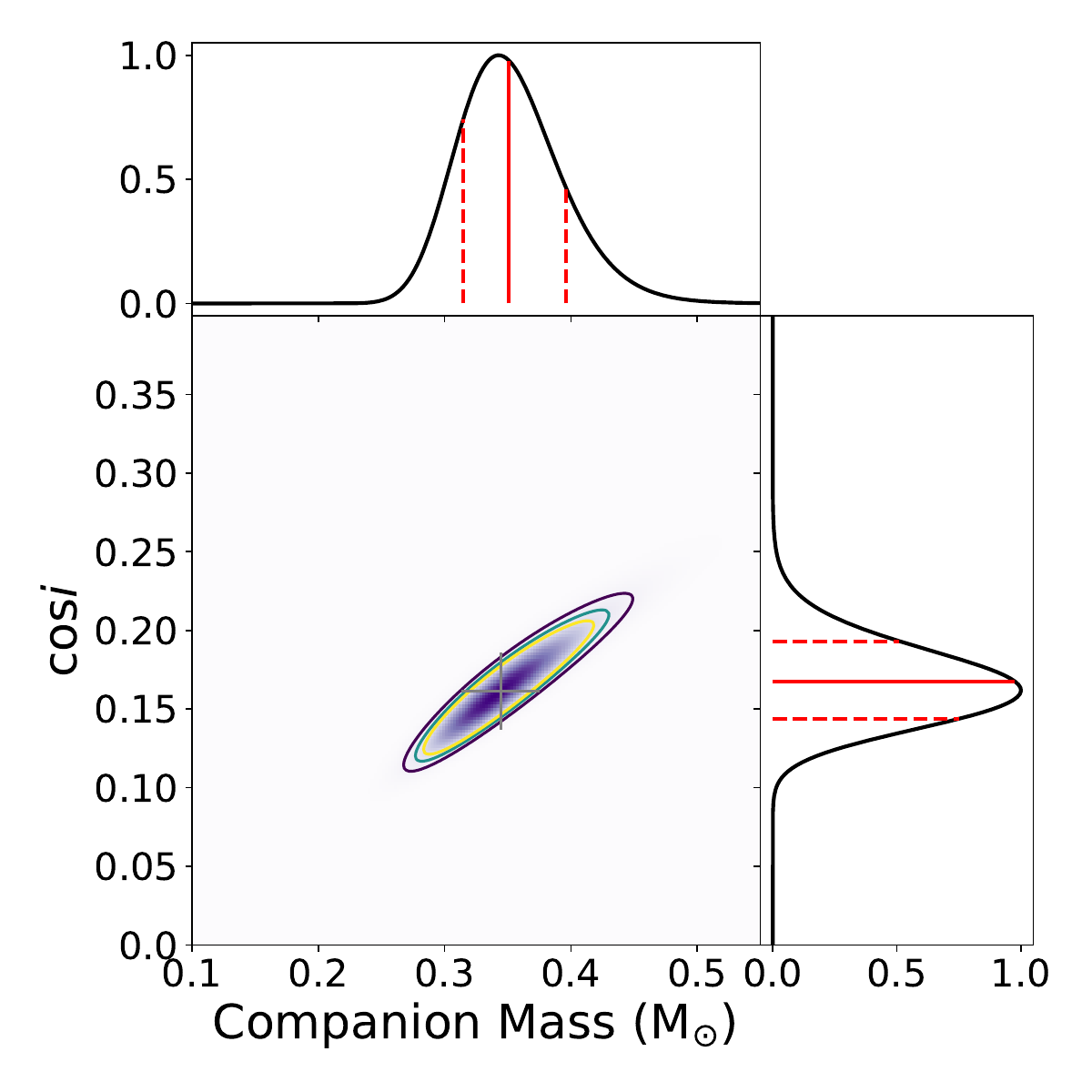}
    \caption{PSR~\jtwentythree\ measurement of companion mass and cosine of inclination ($\cos i$) from Shapiro delay. Center plot is the 2-D posterior pdf in purple for $m_{\rm c}$ and $\cos i$ with 1,2,3-$\sigma$ contours overlaid. The grey cross in the center is the analytically determined 68\% confidence intervals for $m_{\rm c}$ and $\cos i$. Top plot is the marginalized 1-D PDF for $m_{c}$, and right side plot is the marginalized 1-D PDF for $\cos i$. Solid red lines indicate the measurement value, and dotted red lines indicate bounds of the 68\% confidence intervals.  }
    \label{fig:J2302_m2_cosi}
\end{figure}

\begin{figure}
    \centering
    \includegraphics[width=\linewidth]{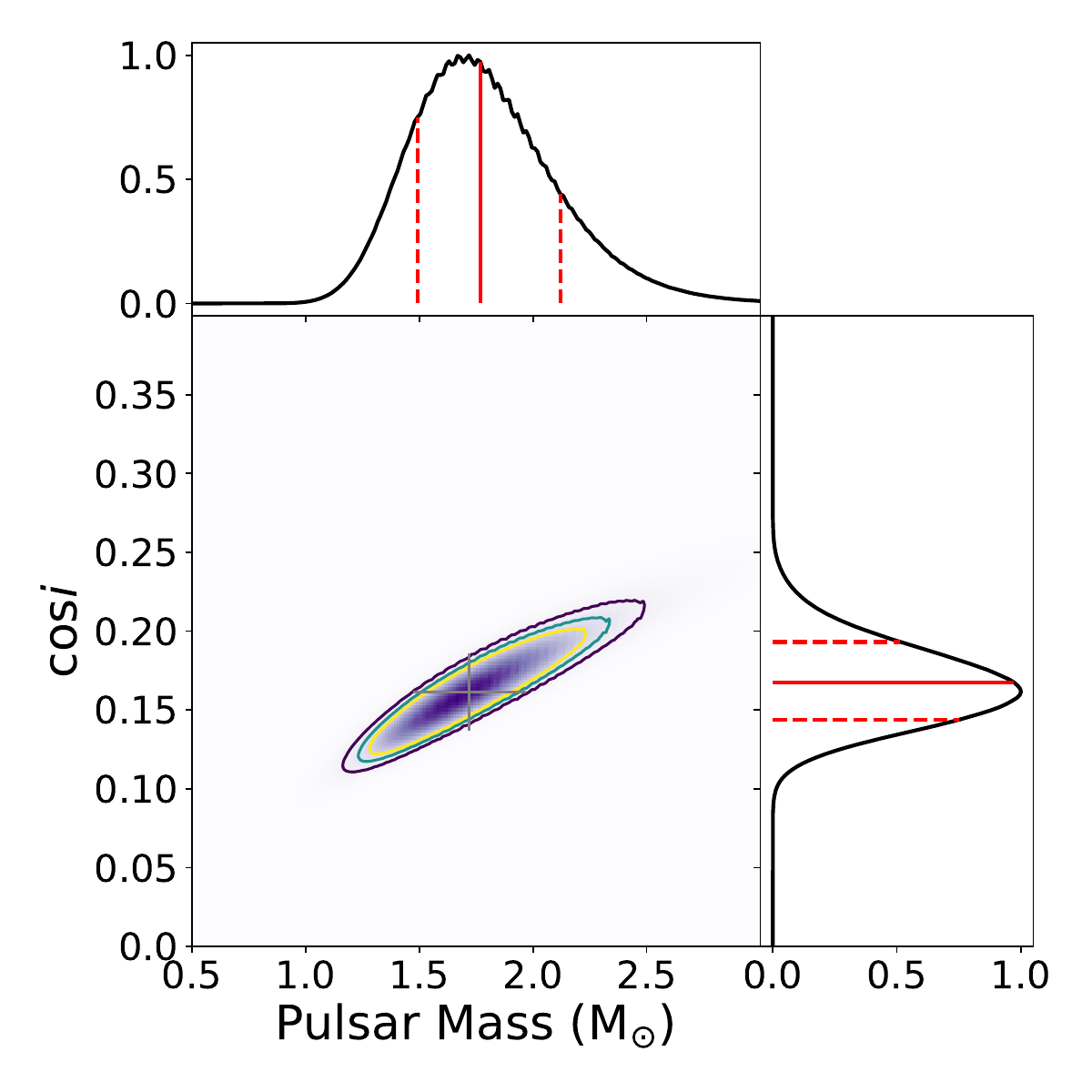}
    \caption{PSR~\jtwentythree\ pulsar mass constraints from Shapiro delay measurement. Center plot is the 2-D posterior pdf in purple for $m_{\rm p}$ and $\cos i$ with 1,2,3-$\sigma$ contours overlaid. The grey cross in the center is the analytically determined 68\% confidence intervals for $m_{\rm p}$ and $\cos i$. Top plot is the marginalized 1-D PDF for $m_{p}$, and right side plot is the marginalized 1-D PDF for $\cos i$. Solid red lines indicate the measurement value, and dotted red lines indicate bounds of the 68\% confidence intervals. Jagged line in the top plot is an artifact of the interpolator used for plotting. }
    \label{fig:J2302_mp_cosi}
\end{figure}

We determined confidence intervals for $m_{\rm c}$, $m_{\rm p}$, and $\cos i$ by marginalizing each of the 2-D PDFs over one variable to get the 1-D PDFs of the corresponding variable (Fig.~\ref{fig:J2302_m2_cosi} and \ref{fig:J2302_mp_cosi}). For $m_{\rm c}$ we got $0.35^{+0.05}_{-0.04} \,M_{\odot}$, $m_{\rm p}$ was $1.8^{+0.3}_{-0.3} \,M_{\odot}$, and $\cos i $ was $0.17^{+0.03}_{-0.02}$ which corresponds to an inclination angle of $80^{\circ+1}_{-2}$. 

Comparing this result to the NG15 Shapiro delay results, we switched the binary model from DD to ELL1H due to the nearly circular eccentricity, and thus used a different parameterization. In the DD binary model, Shapiro delay is parameterized in the timing model by 
$m_{\rm c}$ and $\sin i $ \citep{dd_model_1,dd_model_2}.  
The combined CHIME/NG15 results show significant improvement from the Shapiro-delay estimated mass measurements from \citet{NG15_timing}, which reported an $m_{\rm c}$ of $0.3\pm 0.1 \,M_{\odot}$, $m_{\rm p}$ of $5.2^{+3.0}_{-3.2}\,M_{\odot}$, and system inclination of $78\degree \pm 3.3\degree$. Our results for $m_{\rm c}$ and $i$ have improved precision by a factor of 2 and are consistent within the 1-$\sigma$ error region. Our $m_{\rm p}$ measurement uncertainty is dramatically lower by two orders of magnitude and is well-determined in a more physically motivated range of pulsar masses than the analysis in \citet{NG15_timing}. This improvement is possibly due to a combination of the longer timing baseline having more full orbits, this pulsar has a long binary period of 125 days, and the change to the ELL1H binary model to account for the nearly circular orbit and high system inclination.  

%% Switched binary model because of low eccentricity and passing the Ell1 check: $x$ $e^{4}$ << TRES/ $\sqrt{N_{\rm TOAs}}$      
%asini/c * ecc**4    = 3.29e-06 us TRES / sqrt(# TOAs) = 0.0508 us 
%NG15 comparison - Shapiro delay was determined using the DD binary model with companion mass of 0.3 +/_ 0.1  solar masses and sini of 0.98 +/- 0.01
% NG 15 shapiro delay determined pulsar mass - 5.2 +3.2 / -3.0 

\subsection{Solar wind modeling}\label{subsec:swx}

The solar wind is typically modeled in pulsar timing as a spherical distribution of free electrons that is constant in time. In the timing model it is parameterized with a solar wind electron density ($n_{e}$) parameter and a radial power law index \citep{2006_edwards_hobbs}. DM variations from this model can be calculated as follows: 

\begin{equation}
    DM_{SW} = n_e(1 \text{AU})^{2} \frac{\theta}{r_{e}\sin \theta}
\end{equation}

\begin{figure*}
    \centering
    \includegraphics[width=\linewidth]{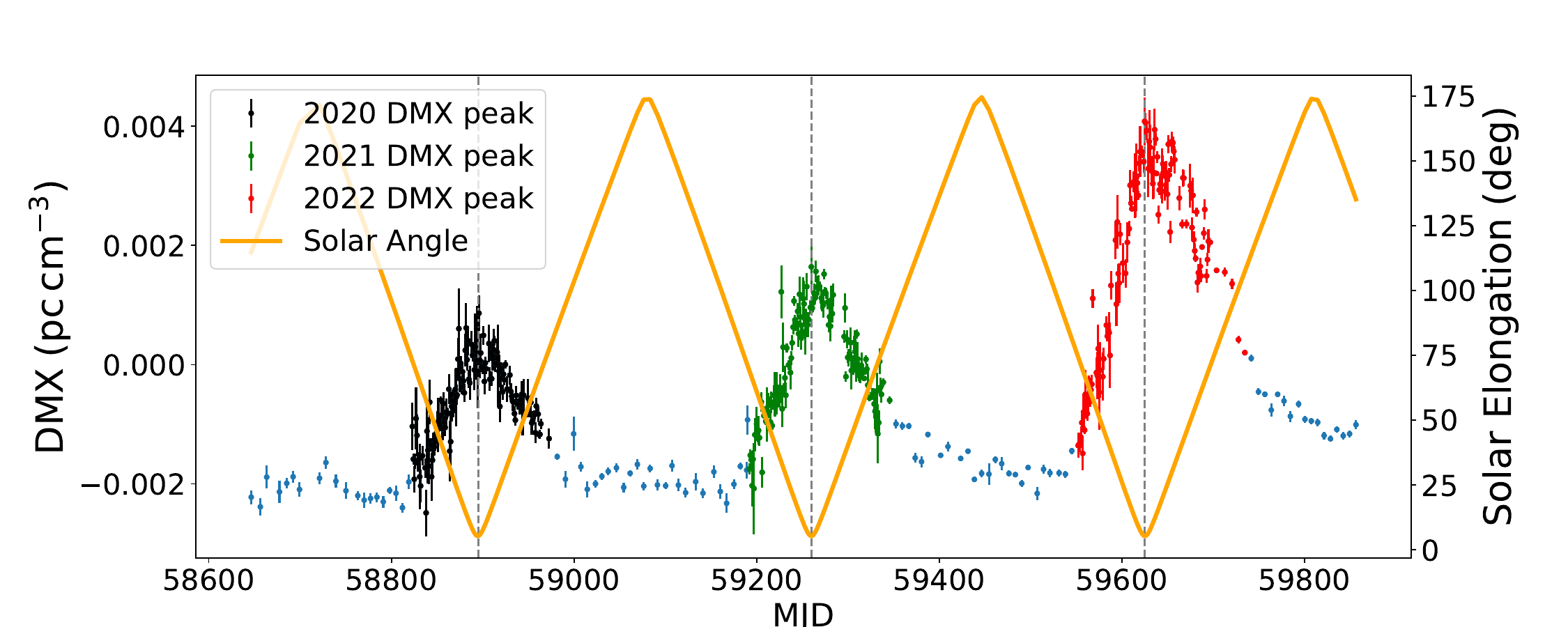}
    \caption{Partial DMX time series for CHIME-era PSR~\jtwentyone\ data. The gold line indicates solar angle and the dotted grey lines represent each annual conjunction. The DMX points corresponding to bins where we detect excess solar wind DM are color coded by year. }
    \label{fig:J2145_sw_peaks}
\end{figure*}

Where $n_e$ is the solar wind electron density at 1 AU, $r_e$ is the Earth-Sun distance, and $\theta$ is the pulsar-Sun-observatory angle. In low ecliptic latitude pulsars, these DM variations can be visible in the DM time series as annual peaks that correspond to a minimized solar angle \citep{2019_Tibruzi}. This effect is particularly well-illustrated in the CHIME era data for PSR~\jtwentyone\ (Fig.~\ref{fig:J2145_sw_peaks}). 

Many previous studies have demonstrated that the radial model is insufficient to precisely model $DM_{SW}$ variations on both long and short timescales \citep{2012_you,2019_Tibruzi,2022_hazboun}. For PSR~\jtwentyone, we noticed that the DMX time series doesn't show constant amplitudes of the $DM_{SW}$ peaks, and that the peaks are not symmetrical, nor consistent with a radial power law fall off. As seen in Figure~\ref{fig:J2145_sw_peaks}, the solar wind peaks have a much shallower slope than we would expect with the radial model. 

We tried several methods of modeling the SW. When using the standard radial model, we found it to be highly covariant with the DMX model, and observed sharp dips in the DMX time series points that corresponded to conjunction. We also tried the \texttt{pint} \texttt{SolarWindDispersionX} model, which is a piece-wise constant representation of the solar wind density parameterized by a maximum solar wind DM at conjunction and a power-law index \citep{2024_PINT}. This model assumes that the excess solar wind DM is zero at opposition. We used a single SWX bin for data prior to MJD 58600 as the detected excess $DM_{SW}$ was relatively low and not significantly different from one year to the next. For data after MJD 58600 we used one SWX bin per year. We found that this model was unable to capture the asymmetry in the SW peaks and still left significant SW-like structure in our DMX time series and WB DM residuals. For our analysis, we found the simplest and most robust method of modeling all the $DM_{SW}$ variations was to use more dense DMX binning for data post MJD 58600. This alternative method still did not entirely eliminate solar wind-like structure from the DM residuals for this source as shown in Figure~\ref{fig:J2145_resids}. One of the largest SW peaks in 2022 also corresponds to a dip in the TOA residuals. In future analyses, the CHIME data for this system could be used to test more complex solar wind models that can better separate ISM contribution to DM vs solar wind contribution.

%Similarly to previous analyses 
% SOlar wind contribution to DM
% It's normally modeled spherical and constant in time but we know it is not. 
\section{Discussion}\label{sec:discussion}

The combination of four years of CHIME data with NG15 data have already massively increased the TOA volume per source. Even for sources like PSR~\jten\ and PSR~\jtwentyone\ that have the longest NG15 timing baseline, there are already enough CHIME TOAs to nearly triple the total TOA volume (Table~\ref{tab:timing_stats}). 
This increase will get compounded in narrowband datasets made with CHIME data as there could be anywhere from several to dozens of TOAs per observations to allow proper frequency coverage. For comparison, in NG15 the narrowband datasets for the sources in this paper had at least 41 times the number of TOAs as the corresponding NG15 WB datasets. As CHIME data is currently being incorporated into multiple upcoming PTA datasets, such as NANOGrav's upcoming 20-year dataset and IPTA's third data release, this sudden explosion of TOAs is likely to create a data volume issue for both single source analyses and PTA analyses.

\subsection{PK parameter improvements}

The fractional uncertainty of PK parameters is expected to decrease with increasing observing baseline. The $\dot{P}_{B}$ measurement for PSR~\jten\ decreased in fractional uncertainty between the 12.5 yr dataset and NG15 by the theoretical expected improvement of $T^{-5/2}$ \citep{1992_Damour_Taylor,12.5yr_wideband,NG15_timing}, where $T$ represents the  observing time in years. Similarly the improvement in fractional uncertainty with the inclusion of four years of CHIME data to NG15 was also $T^{-5/2}$. We also calculated that the fractional uncertainty of $\dot{x}$ for PSR~\jten\ improved by roughly $T^{-2.1}$ between NG15 and our CHIME/NG15 analysis. This parameter was newly detected for this source NG15 so we could not compare to the 12.5 yr dataset. We were able to do a comparison for the $\dot{x}$ measurement for PSR~\jtwentyone, where the 12.5-yr to NG15 improvement was $T^{-2.3}$ and the NG15 to CHIME/NG15 improvement was $T^{-2.2}$. These very similar improvements in fractional uncertainty likely indicate that the higher cadence of the CHIME data is less important for the improvement of PK parameter measurements. The inclusion of CHIME data mainly benefits these PK measurements by significantly increasing the timing baseline.

\subsection{CHIME Frequency coverage and WB TOA precision}

\begin{figure*}
    \centering
    \includegraphics[width=\textwidth]{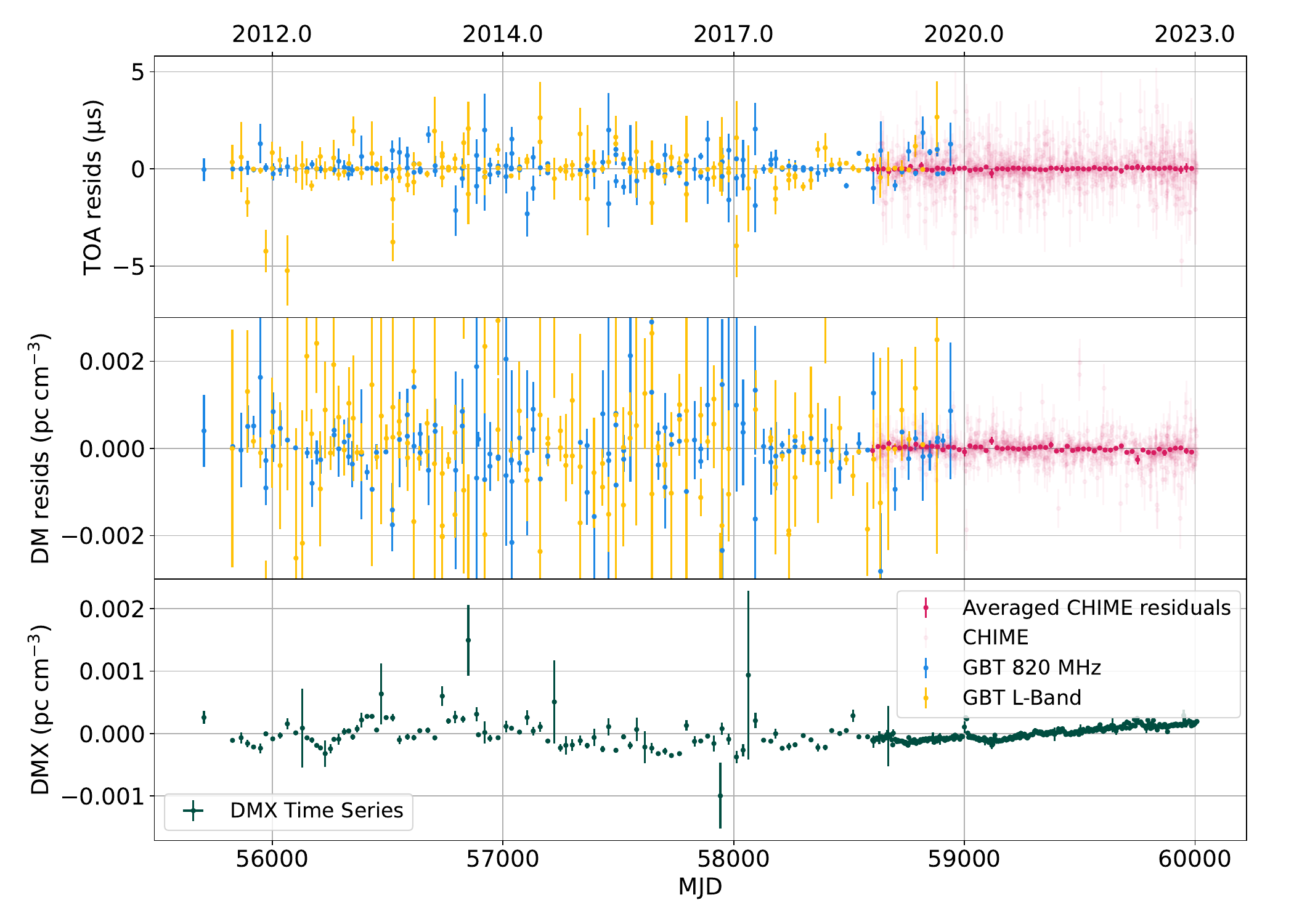}
    \caption{Top plot: TOA residuals for PSR~\jzerosix; Middle plot: WB DM residuals; Bottom plot: DMX time series where vertical bars indicate DMX parameter uncertainty and horizontal bars indicate the width of DMX bins in time. }
    \label{fig:J0645_resids}
\end{figure*}

\begin{figure*}
    \centering
    \includegraphics[width=\textwidth]{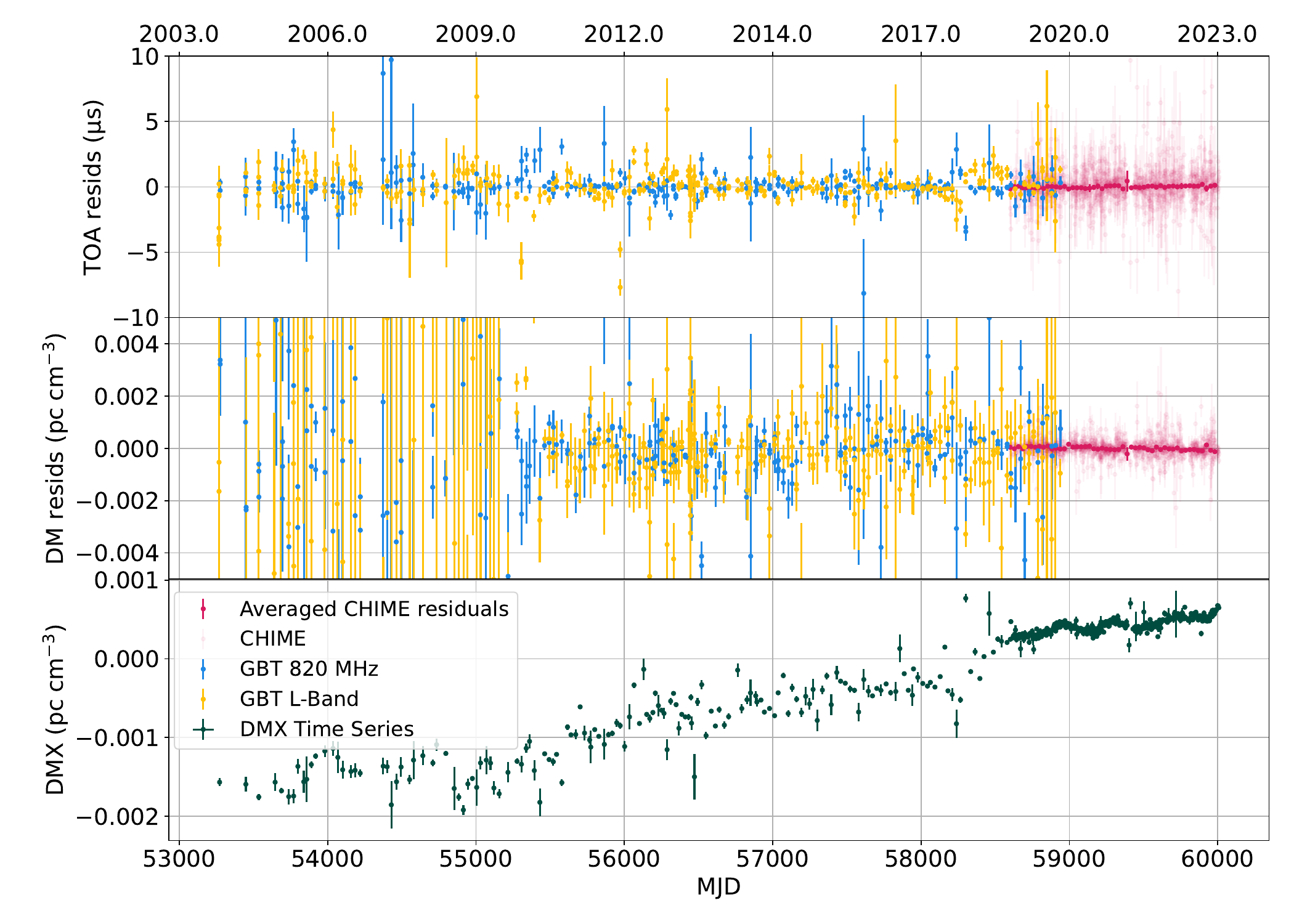}
    \caption{Top plot: TOA residuals for PSR~\jten; Middle plot: WB DM residuals; Bottom plot: DMX time series where vertical bars indicate DMX parameter uncertainty and horizontal bars indicate the width of DMX bins in time. }
    \label{fig:J1012_resids}
\end{figure*}

\begin{figure*}
    \centering
    \includegraphics[width=\textwidth]{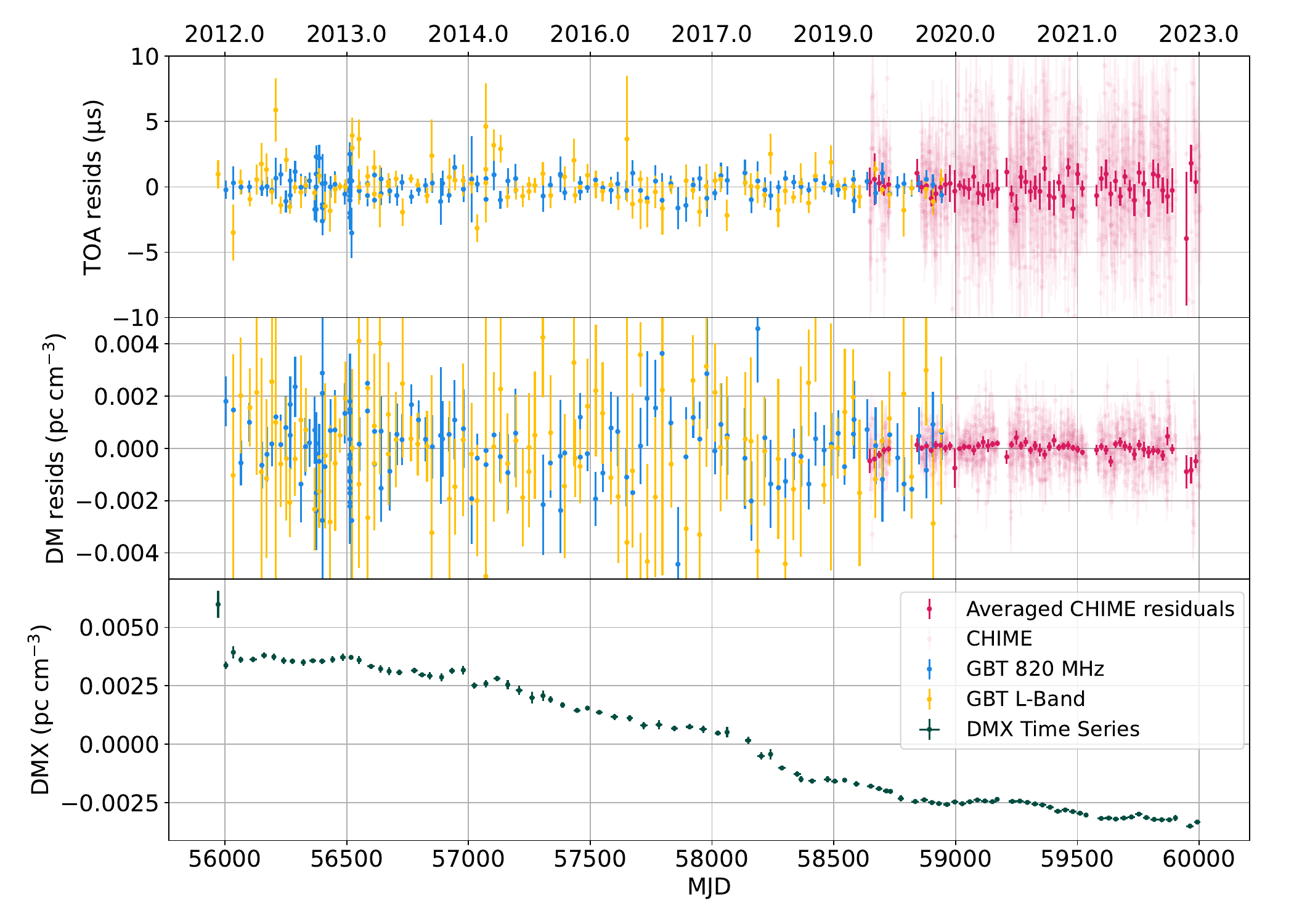}
    \caption{Top plot: timing resids for PSR~\jtwentythree; Middle plot: DM residuals; Bottom plot:DMX time series where vertical bars indicate DMX parameter uncertainty and horizontal bars indicate the width of DMX bins in time. }
    \label{fig:J2302_resids}
\end{figure*}

In nearly all of the timing residual plots it is clear that the CHIME WB TOAs, on average, have larger uncertainties compared to many of the NG15 WB TOAs (Figs~\ref{fig:J0645_resids}, \ref{fig:J1012_resids}, \ref{fig:J2145_resids}, \ref{fig:J2302_resids}). Some of this increase can be attributed to CHIME operating at lower frequencies than nearly all the receivers used for NG15. TOA uncertainty scales as $\nu^{-1/2}$, so we would generally expect high-frequency TOAs to have smaller error bars. The CHIME telescope also has a lower gain, which is inversely proportional to TOA uncertainty, than the GBT \citep{msp_living_review,pulsar_handbook}. 

The corresponding WB DMs are the opposite, with the CHIME WB DMs being much more precise than GUPPI or GASP DMs. This decrease is likely due to CHIME's frequency coverage being larger and covering lower frequencies than any of the GBT receivers used. DM is measured by the delay between pulse arrivals from different frequencies. Lower frequencies will have a much larger lever arm on measuring DM delay, which scales with $\nu^{-2}$. Using PSR~\jzerosix\ as an example, the DM delay across the CHIME bandwidth was 291~ms, whereas the delay across the 820~MHz receiver was 43~ms and the delay across the L-Band receiver would only be 29~ms. ISM propagation effects are a significant source of noise for PTA analyses, so the increase in total observing range due to the inclusion of low-frequency, high precision, CHIME DM data will be important for modeling frequency-dependent DMs \citep{15yr_noise_budget}. The high cadence observing also is beneficial for monitoring short and long timescale DM fluctuations.

Our RFI excision approach to remove bad channels after frequency and time averaging has the potential to cause us to lose a significant percentage of frequency information. When constructing the data portraits we scrunched in frequency to 64 bins and kept 53\% of frequency coverage for PSR~\jzerosix, 68\% for PSR~\jten, 57\% for \jtwentyone, and 40\% for PSR~\jtwentythree. Another compounding factor is that parts of the CHIME observing band are unprotected for radio astronomy as they overlap with frequencies used for cell phones. Thus, CHIME has an extremely variable RFI environment and continuing work is underway to develop more sophisticated automated RFI excision algorithms.  

\section{Conclusions} \label{sec:conclusions}

In this paper we presented timing results from a sample of four MSPs from the NANOGrav 15-yr dataset that have been combined with four years of daily cadence CHIME data. This is one of the first large scale efforts to include CHIME data in PTA datasets in the wideband paradigm. We summarize our major results and conclusions below.

We report significant Shapiro-delay estimated mass constraints for PSR~\jtwentythree. We measured a system inclination of ${80\degree}^{+1}_{-2}$, a companion mass of $0.35^{+0.05}_{-0.04}\,M_{\odot}$, and a pulsar mass of $1.8^{+0.3}_{-0.3}\,M_{\odot}$. These measurements have not only significantly improved from the reported NG15 values, but also from those reported on optical analyses of the white dwarf companion \citep{2018_Kirichenko}.

Using the proper motion and galactic acceleration derived corrections to the measured spin down of PSR~\jzerosix\ we get an intrinsic spindown of $4.5(12) \times 10^{-21} \text{\,s\,s}^{-1}$. The significance of astrometric parameter measurements for PSR~\jzerosix\ improved slightly with the inclusion of CHIME data. 

We report improved pulsar and companion mass constraints for PSR~\jten\ and PSR~\jtwentyone\ using a combination of improved PK parameter measurements and VLBI results. Measurements of post-Keplerian parameters $\dot{P}_{B}$ and $\dot{x}$ improved with the inclusion of a longer timing baseline, but don't appear significantly impacted by the higher cadence of observing. The use of CHIME data to produce more precise $\dot{P}_{B}$ measurements combined with improved pulsar distances can also benefit studies of local galactic potential gradients through improved pulsar binary line of sight acceleration estimates \citep{Moran_2023,Moran_2024}.

As PTAs become more sensitive, with both longer timing datasets and increased observing bandwidths, the potential for future detection of continuous waves from individual supermassive black hole binaries will grow. The inclusion of CHIME data in PTA datasets is particularly important for studying interstellar propagation effects which can significantly impact PTA sensitivity. In future NANOGrav datasets, CHIME data will help mitigate the loss of low-frequency observing due to the collapse of the Arecibo Telescope. CHIME WB DM data also presents a unique laboratory for studying the effect of the solar wind on pulsar timing. The more precise WB DM measurements we get from the low-frequency, high cadence CHIME data are extremely useful for illustrating short time scale ISM fluctuations, particularly DM variability from the solar wind. This was particularly well-demonstrated in the DM data for PSR~\jtwentyone. In future analyses CHIME data could be used to build more precise global solar wind models.

%What do I want to say: CHIME will mitigate the loss of Arecibo in future PTA datasets. 

\clearpage

This work has been carried out as part of the NANOGrav collaboration, which receives support
from the National Science Foundation (NSF) Physics Frontiers Center award numbers 1430284 and
2020265.
G.Y.A was supported in part by NASA under Award Nos. RFP22.1-0 and RFP23\_1-0 through the Wisconsin Space Grant Consortium Graduate and Professional Research Fellowship. 
P.R.B.\ is supported by the Science and Technology Facilities Council, grant number ST/W000946/1.
H.T.C.\ acknowledges funding from the U.S. Naval Research Laboratory.
Pulsar research at UBC is supported by an NSERC Discovery Grant and by CIFAR.
K.C.\ is supported by a UBC Four Year Fellowship (6456).
F.A.D is supported by an NRAO Jansky fellowship.
M.E.D.\ acknowledges support from the Naval Research Laboratory by NASA under contract S-15633Y.
T.D.\ and M.T.L.\ received support by an NSF Astronomy and Astrophysics Grant (AAG) award number 2009468 during this work.
E.C.F.\ is supported by NASA under award number 80GSFC24M0006.
D.C.G.\ is supported by NSF Astronomy and Astrophysics Grant (AAG) award \#2406919.
D.R.L.\ and M.A.M.\ are supported by NSF \#1458952.
M.A.M.\ is supported by NSF \#2009425.
The Dunlap Institute is funded by an endowment established by the David Dunlap family and the University of Toronto.
T.T.P.\ acknowledges support from the Extragalactic Astrophysics Research Group at E\"{o}tv\"{o}s Lor\'{a}nd University, funded by the E\"{o}tv\"{o}s Lor\'{a}nd Research Network (ELKH), which was used during the development of this research.
H.A.R.\ is supported by NSF Partnerships for Research and Education in Physics (PREP) award No.\ 2216793.
S.M.R.\ and I.H.S.\ are CIFAR Fellows.
Portions of this work performed at NRL were supported by ONR 6.1 basic research funding.

The National Radio Astronomy Observatory and Green Bank Observatory are facilities of the National Science Foundation operated under cooperative agreement by Associated Universities, Inc.

We acknowledge that CHIME is located on the traditional, ancestral, and unceded territory of the Syilx/Okanagan people. We are grateful to the staff of the Dominion Radio Astrophysical Observatory, which is operated by the National Research Council of Canada.  CHIME is funded by a grant from the Canada Foundation for Innovation (CFI) 2012 Leading Edge Fund (Project 31170) and by contributions from the provinces of British Columbia, Qu\'{e}bec and Ontario. The CHIME/FRB Project, which enabled development in common with the CHIME/Pulsar instrument, is funded by a grant from the CFI 2015 Innovation Fund (Project 33213) and by contributions from the provinces of British Columbia and Qu\'{e}bec, and by the Dunlap Institute for Astronomy and Astrophysics at the University of Toronto. Additional support was provided by the Canadian Institute for Advanced Research (CIFAR), McGill University and the McGill Space Institute thanks to the Trottier Family Foundation, and the University of British Columbia. The CHIME/Pulsar instrument hardware was funded by NSERC RTI-1 grant EQPEQ 458893-2014.

This research was enabled in part by support provided by the BC Digital Research Infrastructure Group and the Digital Research Alliance of Canada (alliancecan.ca).

\textit{Facilities}: Green Bank Observatory (GBO), Dominion Radio Astrophysical Observatory

\textit{Software}: \verb+PINT+ \citep{2021_PINT,2024_PINT}, \verb+PulsePortraiture+ \citep{pulse_portraiture}, \verb+enterprise+ \citep{enterprise_software}, Matplotlib \citep{Hunter:2007}, Numpy \citep{harris2020array}, This work made use of Astropy:\footnote{http://www.astropy.org} a community-developed core Python package and an ecosystem of tools and resources for astronomy \citep{astropy:2013, astropy:2018, astropy:2022}.

\appendix
\section{The \texttt{chimera} pipeline} 
\label{sec:chimera}

We developed the \texttt{chimera} pipeline\footnote{Available at \url{https://doi.org/10.5281/zenodo.18458818}.} to measure wideband TOAs from the partially folded profiles in the \texttt{Timer} format generated by the CHIME/Pulsar backend \citep{chimera}.
\texttt{chimera} is written in Python and provides a command-line script named \texttt{chimerawb}\footnote{\texttt{chimera} is written with a structure similar to and reuses code from the \texttt{pinta} pipeline of the Indian Pulsar Timing Array \citep{pinta_inpta}.
}.
The \texttt{chimerawb} command takes an input directory containing \texttt{Timer} archives \citep{dspsr},
a configuration file in the \texttt{JSON} format, optionally a meta file containing the list of input archives to be processed, and an output directory as input, and writes a \texttt{tim} file containing wideband TOAs in the IPTA format into the output directory.
The configuration file contains information about the pulsar name, fiducial DM, template portrait, number of subintegrations/channels for time/frequency scrunching, and a list of bad channels for pulsar-specific RFI flagging.
The template portrait is generated manually beforehand using the \texttt{ppspline} command of \texttt{PulsePortraiture} from an epoch-averaged profile obtained from multiple observations, since individual CHIME observations may have low S/N \citep{pulse_portraiture}; see section~\ref{subsec:wb_timing}. 
The TOAs are generated using the \texttt{pptoa} command of \texttt{PulsePortraiture} \citep{pulse_portraiture}.

The specific operations performed by the \texttt{chimerawb} command are listed below:
\begin{enumerate}
    \item For each \texttt{Timer} file in input directory:
    \begin{enumerate}
        \item Perform automated RFI flagging.
        \item Convert from \texttt{Timer} to \texttt{PSRFITS} format \citep{psrfits_psrchive}.
        \item Scrunch in time and frequency.
        \item Perform pulsar-specific RFI flagging using the given list of bad channels.
    \end{enumerate}
    \item Generate wideband TOAs and save them as a \texttt{tim} file.  
\end{enumerate}

An example invocation of \texttt{chimerawb} is given below:
\begin{verbatim}
    $ chimerawb -i J1234+5678_data/ -m J1234+5678.meta  -c J1234+5678_cfg.json   \
                -o J1234+5678_output/
\end{verbatim}
Here, the \texttt{-i} argument is the input directory, the \texttt{-m} option provides the meta file containing the list of archives to be processed, the \texttt{-c} argument provides the configuration file, and the \texttt{-o} argument provides the output directory.

\bibliography{references.bib}
\bibliographystyle{aasjournal}

\end{document}